\documentclass[%
 reprint,
 amsmath,amssymb,
 aps,
]{revtex4-2}

\usepackage{graphicx}
\usepackage{dcolumn}
\usepackage{bm}
\usepackage{xcolor}
\usepackage{amsmath,amssymb}

\usepackage{microtype}
\usepackage{nonfloat}
\usepackage{graphicx}
\usepackage{subfigure}
\usepackage{booktabs} 

\usepackage{amsfonts}
\usepackage{color}

\usepackage{multirow}

\usepackage[colorlinks,linkcolor=black,citecolor=blue,urlcolor=blue ]{hyperref}

\def\Im{{\bf I}}
\def\E{{\mathbb E}}

\begin{document}

\preprint{APS/123-QED}

\title{Cosmological Parameter Estimation and Inference using Deep Summaries}

\author{Janis~Fluri}
 \email{janis.fluri@phys.ethz.ch}
\affiliation{Institute of Particle Physics and Astrophysics\\
             Department of Physics, ETH Zurich \\
             Switzerland
            }
\author{Aurelien~Lucchi}
\affiliation{Data Analytics Lab \\ 
             Department of Computer Science , ETH Zurich \\
             Switzerland
            }
\author{Tomasz~Kacprzak}%
\affiliation{Institute of Particle Physics and Astrophysics\\
             Department of Physics, ETH Zurich \\
             Switzerland
            }
\author{Alexandre Refregier}
\affiliation{Institute of Particle Physics and Astrophysics\\
             Department of Physics, ETH Zurich \\
             Switzerland
            }
\author{Thomas~Hofmann}
 \affiliation{Data Analytics Lab \\ 
             Department of Computer Science , ETH Zurich \\
             Switzerland
            }

\date{\today}

\begin{abstract}
The ability to obtain reliable point estimates of model parameters is of crucial importance in many fields of physics. This is often a difficult task given that the observed data can have a very high number of dimensions. In order to address this problem, we propose a novel approach to construct parameter estimators with a quantifiable bias using an order expansion of highly compressed deep summary statistics of the observed data. These summary statistics are learned automatically using an information maximising loss. Given an observation, we further show how one can use the constructed estimators to obtain approximate Bayes computation (ABC) posterior estimates and their corresponding uncertainties that can be used for parameter inference using Gaussian process regression even if the likelihood is not tractable. We validate our method with an application to the problem of cosmological parameter inference of weak lensing mass maps. We show in that case that the constructed estimators are unbiased and have an almost optimal variance, while the posterior distribution obtained with the Gaussian process regression is close to the true posterior and performs better or equally well than comparable methods. 
\end{abstract}

\maketitle

\tableofcontents

\section{Introduction}

Parameter estimation and inference are common problems in many fields of physics. In parameter estimation, one tries to construct unbiased point estimators for model parameters and in parameter inference one is interested in the posterior distribution of the underlying model parameters, given a set of observations. These two problems are often difficult to solve in real-world scenarios for multiple reasons. There are usually only very few relevant parameters in physical models that are used to describe highly complex systems whose interactions are governed by physical laws. Due to the nature of these complex interactions it is rarely possible to directly observe the underlying model parameters, but rather one has to extract them from high dimensional data. A prime example for this is the field of cosmology or, more specifically, weak graviational lensing (see e.g. \cite{WLreview} for a review). The currently most accepted $\Lambda$CDM (dark energy, cold dark matter) cosmological model contains only five relevant parameters that have to be extracted from the shape measurements of millions of galaxies. The relationship between the model parameters and the observed data is highly non-linear and not analytically tractable. A practical way to extract the desired parameters is via complex models and large cosmological simulations that can require millions of CPU and GPU hours to run \citep{Potter2017}. Given the large amount of data produced by these simulations, a standing problem is the compression of the observed data, such that one can optimally constrain the cosmological parameters. Frequently used methods for the data compression include MOPED and (nuisance hardened) score compression \cite{compression1,compression2,compression3}. However, the recent advances in machine learning offer new possibilities in the field of cosmology. Especially deep neural networks, have gained a lot of attention \citep{Charnock2018,ribli2019weak} and have been used in particular for parameter estimation \citep{Herbel2018} and inference \citep{Fluri2019,netinference2020}.

An important problem in the construction of parameter estimators is the quantification of their bias. For deep neural networks there has been extensive work concerning uncertainty estimation \citep{Gal2016}. Common approaches include drop out \citep{Gal2016b} and Bayes by Backprop \citep{Charles2015}. These methods help to quantify the intrinsic model uncertainties, however, they do not quantify potential biases of the models. In \cite{Chernozhukov2018} they are able to construct unbiased estimators, but the presented method is not applicable for general non-linear relationships between model parameters and observations that are considered in this work.

In parameter inference one tries to evaluate the posterior distribution and consequently confidence intervals of the underlying model parameters given a set of observations. If the posterior distribution can be evaluated analytically, one can resort to classic methods such as Markov chain Monte Carlo (MCMC) to obtain the desired confidence intervals. However, even if the posterior distribution is known, it might be unfeasible to perform a MCMC run if the likelihood requires the evaluation of expensive simulators. Approaches using Gaussian process regression have been developed \citep{Aretha2019,Pelle2020} in order to reduce the number of required likelihood evaluations and will be discussed in more detail below. Such schemes can potentially reduce the number of necessary likelihood evaluations by orders of magnitude. However, they are not applicable if the likelihood is either unknown or can't be evaluated. In these cases one can resort to likelihood free inference (LFI) methods such as approximate Bayesian computation (ABC) \citep{Fan2018}. ABC methods make use of a discrepancy measure between observed and simulated data to estimate the posterior distribution. Given a sufficient amount simulations it is theoretically possible to obtain the true posterior via this measure and a kernel function. However, ABC methods, while accurate and stable, again require a large number of simulations that might be unfeasible to run in a reasonable time. 

Combinations of Gaussian process regression and ABC have also been proposed, e.g. to emulate the discrepancy measure between simulations \citep{Jarven2017}. This emulation can greatly reduce the required number of simulations, as the posterior can be estimated directly through the emulated discrepancy measure. Further, \cite{Jarven2019} analysed Gaussian process regression of noisy log-likelihood estimates. This was primarily done in a synthetic likelihood setting, where the unknown likelihood is approximated by a Gaussian distribution. Using simulations, it is possible to estimate the Gaussian approximation of the likelihood, as well as, its uncertainty. This approximation can then be emulated with Gaussian processes to obtain the posterior. In \cite{Jarven2019} they then focus on the  uncertainty estimation of the emulated posterior and acquisition strategies for new simulations that optimally improve improve the posterior estimate. 

There are also combinations of ABC and deep neural networks, for example learning summary statistics for ABC inference has been proposed in \cite{sum1,sum2,sum3}, e.g. by training the networks to predict the model parameters directly from simulated data. Further, a method to generate globally sufficient summaries of arbitrary data using mutual information has been developed in \cite{sum4}. However, these approaches usually require on-the-fly simulations and do not use the generated summaries for parameter estimation.

Likelihood-free inference using neural networks has been used in \cite{lfi1,lfi2,lfi3,Alsing2019}. These approaches mostly aim to learn the distribution of the (compressed) data that can be used to obtain the posterior distribution of the underlying model parameters and are not considering a fixed grid of pre-computed simulations. An example of such an approach using neural density estimation to cosmological data can be found in \cite{cosmo_application}. Lastly, in \cite{similar_paper} they examine the compression of weak lensing data similiar to the approach in this work, the key difference being that they have access to on-the-fly simulations and use neural density estimation for the inference instead of our approach that is explained below.

\begin{figure}[t]
    \centering
    \includegraphics[width=0.45\textwidth]{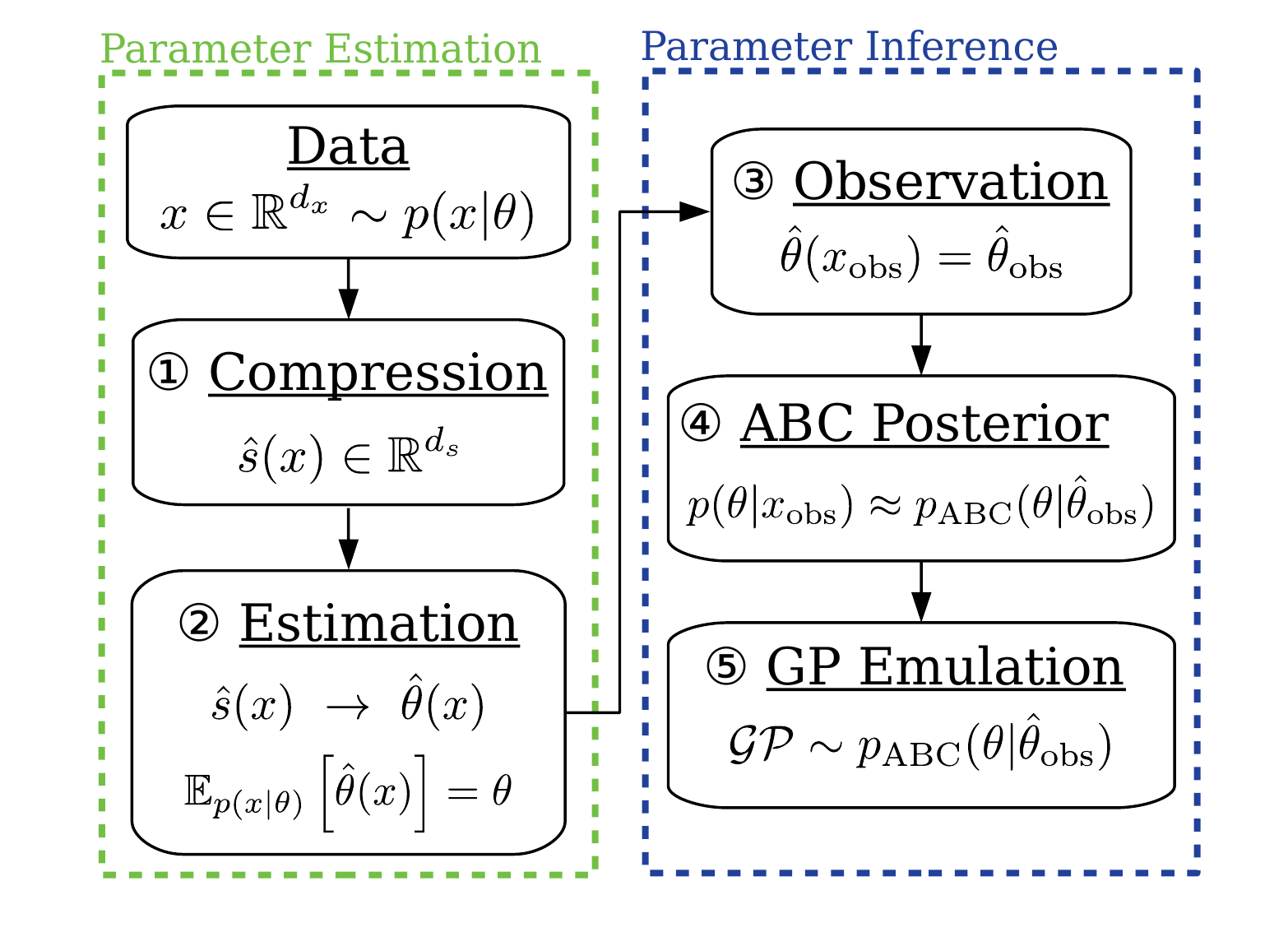}
    \caption{Overview of the presented approach that consists of two parts. In the first part we construct parameter estimators $\hat{\theta}$ using the highly compressed summaries $\hat{s}(x)$ of the simulated data $x$. And in the second part we use these estimators to calculate the ABC posterior $p_\mathrm{ABC}$ which is then emulated with a Gaussian process regression to find the confidence regions of the model parameters. \label{fig:overview}}
\end{figure}

In this work, we consider a scenario that is especially common in weak graviational lensing, where we only have access to a fixed grid of simulations generated prior to the analysis, additional simulations are therefore not obtainable, making it impossible to use acquisition strategies like the ones discussed in \cite{Jarven2019}. The problem we address is to construct estimators of the underlying model parameters and to calculate confidence intervals of these parameters given a set of observations without having access to the likelihood function. An overview of the general approach is shown in Figure~\ref{fig:overview} where we distinguish between two 
parts of the pipeline: i) parameter estimation, where we construct an estimator of the model parameters and ii) parameter inference where we find confidence regions for the model parameters using the estimator constructed in i). To achieve this, we start by relaxing the prior assumptions about the distribution of the compressed data made in \cite{Charnock2018} on their loss function.  Next, we develop a method to construct estimators of the underlying model parameters with quantifiable biases using the compressed simulations. Finally, we demonstrate how one can utilize the constructed estimators to calculate ABC log-posterior estimates and their corresponding uncertainties, which can in turn be used  for parameter inference using Gaussian process regression.
Each of these aspects is important to run the pipeline as a whole in a realistic setting and we validate our findings using a cosmological model that is comparable to modern weak lensing surveys (e.g. \cite{Dominik2020}). Finally, the code used in this work is available on \texttt{github}\footnote{\url{https://github.com/jafluri/cosmo_estimators}}.

\section{Problem Description}
\label{sec:problem}

We start with a detailed description of the two problems of parameter estimation and inference. In parameter estimation, the goal is to estimate the unobserved, usually low dimensional, model parameter $\theta \in \Theta \subseteq \mathbb{R}^{d_\theta}$ from observed, high dimensional data $x \in \mathcal{X} \subseteq \mathbb{R}^{d_x}$. The constructed estimator $\hat{\theta}(x)$ should be unbiased, i.e. 
\begin{equation}
\E_{p(x\vert\theta)}[\hat{\theta}(x)] \equiv \E[\hat{\theta}(x)\vert \theta] \equiv \int_\mathcal{X}\hat{\theta}(x)p(x\vert\theta)\mathrm{d}x = \theta
\end{equation} 
 and minimize the variance around the true value, i.e.
\begin{equation}
\mathrm{Var}[\hat{\theta}_i](\theta) =  \E_{p(x\vert\theta)}[(\hat{\theta}_i(x) - \theta)_i^2].
\end{equation} 
Such an estimator does not always exist. For example, if we have a family of distributions which fulfills $p(x\vert\theta) = p(x\vert\theta')$ for $\theta \neq \theta'$, it is impossible to construct an unbiased estimator $\hat{\theta}(x)$ for all $\theta \in \Theta$. We therefore only consider distributions $p(x\vert\theta)$ that allow the construction of such an estimator.

In parameter inference the goal is to calculate the posterior distribution $p(\theta \vert x)$ of the model parameter given an observation $x$ and subsequently to find confidence intervals for the model parameter. In general this is difficult, because the likelihood $p(x\vert\theta)$ is often unknown. However, we assume that there exists a computationally expensive simulator, which can produce samples of the observed space $x \sim p(x\vert\theta)$ for any possible model parameter $\theta$. Throughout this work, we will consider an experiment design where the computationally expensive simulator was run on a set of $n_\theta$ different model parameters and $n_x$ samples were drawn from the distribution $p(x\vert\theta_\alpha)$ for each parameter $\theta_\alpha$ in the set. We will use Greek indices to indicate parameters contained in a set (e.g. $\theta_\alpha \in \mathbb{R}^{d_\theta}$) and Latin indices (e.g. $\theta_i \in \mathbb{R}$) to refer to individual elements of vectors and matrices.

\section{Parameter Estimation}
\label{sec:estimation}

We start with a description of the first part of the pipeline shown in Figure~\ref{fig:overview}, where our goal is to construct an estimator of the underlying model parameters. A simple way to construct such an estimator $\hat{\theta}(x)$ using the generated simulations is by doing regression. One can train a neural network $h(x,w)$, parameterized by weights $w \in \mathbb{R}^{d_w}$ to predict the underlying model parameter $\theta$ in a supervised fashion using, for example, the mean squared error (MSE) over mini-batches of size $N_b$,
\begin{equation}
L(w) = \frac{1}{N_b}\sum_{\alpha=1}^{N_b} \left\Vert h(x_\alpha, w) - \theta_\alpha^\mathrm{True}\right\Vert^2, \label{eq:square_loss}
\end{equation}
where $\theta_\alpha^\mathrm{True}$ is the true parameter of the randomly sampled simulation $x_\alpha$. However, using this approach, there is no guarantee that the trained network $h(x,w)$ is an unbiased estimator of the model parameter. In fact, the resulting estimator $\hat{\theta}(x) = h(x,w^*)$, where $w^*$ are the trained weights obtained by minimizing the loss function of equation \eqref{eq:square_loss}, will approximate the expected value $\E_{p(\theta\vert x)}[\theta]$ \citep{sum2} and can be arbitrarily biased. 
In our setting with a fixed grid of simulations an additional source of bias can be introduced by the choice of simulated parameters $\theta_\alpha$. By choosing a fixed grid of parameters $\theta_\alpha$ one implicitly places a prior over the parameters the network tries to estimate. In \cite{ribli2019weak} and \cite{Navaro2020} it was shown that the network can learn this prior, potentially leading to large biases in the predictions of the network in areas where there is a lower parameter sampling density. 

To alleviate this issues, we propose to split the problem into two parts. We start by training a network to produce informative, low-dimensional summary statistics $\hat{s}(x) = h(x,w) \in \mathbb{R}^{d_s}$, with $d_\theta \leq d_s \ll d_x$. In a second step, we use these summary statistics to construct an estimator $\hat{\theta}(x) \equiv \hat{\theta}(\hat{s}(x))$.

\subsection{Obtaining Informative Summaries}

As mentioned earlier, we face the problem of producing a compact representation of the data, which we name \emph{summary} (corresponding to step \textcircled{\scriptsize{1}} in Figure~\ref{fig:overview}). To achieve this, \cite{Charnock2018} proposed to optimize the Cram\'er-Rao bound that expresses a lower bound on the variance of an estimator as a function of the Fisher information, starting with the assumption that the summary statistics generated by a neural network follow a Gaussian distribution. We will next show that this assumption is not necessary.  Formally, the Fisher Information Matrix (FIM) is defined as 
\begin{equation}
\Im_{ij}(\theta) = \E_{p(x\vert\theta)}\left[\frac{\partial\log(p(x\vert\theta)}{\partial \theta_i}\frac{\partial\log(p(x\vert\theta)}{\partial \theta_j} \right],\label{eq:fisher}
\end{equation}
where $\theta_i$ denotes the $i$-th entry of the underlying model parameter $\theta \in \mathbb{R}^{d_\theta}$. The Fisher information matrix is used in the Cram\'er-Rao bound, which bounds the covariance matrix of any summary $\hat{s}(x)$ as
\begin{equation}
\mathrm{Cov}_\theta(\hat{s}) \succcurlyeq \frac{\partial \Psi_\theta(\hat{s})}{\partial \theta}\Im(\theta)^{-1}\frac{\partial \Psi_\theta(\hat{s})}{\partial \theta}^T, \label{eq:cramer}
\end{equation}
where the derivative with respect to $\theta$ is understood as a Jacobian matrix, $A \succcurlyeq B$ indicates that $A - B$ is positive definite and we defined
\begin{equation}
\Psi_\theta(\hat{s}) = \E_{p(x\vert\theta)}\left[\hat{s}(x)\right].
\end{equation}
Note, that if the summary $\hat{s}(x)$ is itself an unbiased estimator of $\theta$, the Jacobian becomes the identity and the covariance matrix is bounded directly by the Fisher information matrix. A summary that achieves equality is called sufficient, if its Jacobian with respect to the underlying model parameter has full rank \cite{schervish2012theory}. This type of sufficiency also implies Bayesian sufficiency, i.e.
\begin{equation}
p(\theta\vert x) = p(\theta\vert \hat{s}(x)),
\end{equation}
which means that the summary retains all relevant information about the model parameters. Similarly to the work of \cite{Blum2010} we will only consider summary statistics having the same dimensionality as the model parameter $d_s = d_\theta$. However, in the appendix \ref{ap:high_dim} we explain how the same approach can be used for summaries with $d_s > d_\theta$. Assuming that the Jacobian is invertible, one can transform equation \eqref{eq:cramer} into
\begin{align}
-\log\det(\Im(\theta)) \leq\ &\log\det(\mathrm{Cov}_\theta(\hat{s})) \label{eq:loss_1} \\
                           &- 2\log\left\vert\det\left(\frac{\partial \Psi_\theta(\hat{s})}{\partial \theta}\right)\right\vert. \nonumber
\end{align}
The right side of this inequality can be used as a loss function to construct informative summaries. Note that this loss is equivalent to the log-determinant of the inverse Fisher matrix used in \cite{Charnock2018}, without relying on the assumption that the summary follows a Gaussian likelihood, showing the generality of the approach. 

Estimating the covariance matrix of the summaries in equation \eqref{eq:loss_1} is straightforward. The simulator, however, is generally not differentiable, preventing the exact calculation of the Jacobian. As in \cite{Charnock2018}, we  calculate the Jacobian via finite differences. One starts by generating a set of simulations using a fiducial parameter\footnote{The term ``fiducial parameter'' refers to a reference parameter typically located in a region of interest and based on a previous analysis.} $\theta_\mathrm{fid}$ and small perturbations of it $\theta_\mathrm{fid}^\pm = \theta_\mathrm{fid} \pm \Delta\theta^\pm$, resulting in $2d + 1$ parameters. Using this set of parameters one can approximate the Jacobian as
\begin{equation}
 \left(\frac{\partial \Psi_\theta(\hat{s})}{\partial \theta}\right)_{ij} \approx \frac{\Psi_{\theta_\mathrm{fid} + \Delta\theta^+}(\hat{s})_j - \Psi_{\theta_\mathrm{fid} - \Delta\theta^-}(\hat{s})_j}{\Delta\theta^+_i - \Delta\theta^-_i},
\end{equation} 
where the expectation $\Psi_\theta(\hat{s})$ is empirically calculated over multiple samples from $p(x\vert\theta)$. This reduces the variance at the price of an increased computational effort spent on a single step during the training. However, training the network only around a single (or a few) fiducial points leads to the advantage that the remaining datapoints can be used for testing and validation. It also reduces the overall training time compared to simple regression networks as one does not have to cycle through the entire data set of simulations. 

\subsubsection{Regularization} 
The minimization of the righthand side of inequality \eqref{eq:loss_1} could be numerically unstable because the loss allows the entries of the Jacobian or the summary statistics itself to become arbitrary large or small. The reason for this is that there are infinitely many summary statistics that have the same Fisher information. In fact, any one-to-one transformation of summary statistics preserves the Fisher information, showing the degeneracy of the possible solutions. To cope with this issue we add a regularization term to the loss
\begin{equation}
L_\mathrm{regu}(w) = \lambda \left\Vert \frac{\Psi_\theta(\hat{s})}{\partial \theta} - \mathbb{I}_{d_\theta} \right\Vert^2,
\label{eq:regu_loss_3}
\end{equation}
where $\lambda$ is a hyper-parameter, $\mathbb{I}_d$ represents the $d$-dimensional identity matrix and we use the Frobenius norm as matrix norm. Additionally one can add a regularization term proportional to the length of the summary vectors $\Vert \hat{s}(x)\Vert$ to avoid arbitrarily large summaries. However, we never observed any sort of over- or underflow in the summary statistics during our experiments and did not use this type of regularization.

\subsubsection{Adding Expert Knowledge} 

We briefly mention that the approach described above allows for an easy incorporation of expert knowledge into the summaries. Consider the scenario where a set of summaries $(\hat{s}_1(x), \dots, \hat{s}_n(x))$ generated using expert knowledge is available. One can simply incorporate this additional information into the model by concatenating the summaries to the fully connected layers.  Another alternative is to perform another round of compression by training a different neural network to compress the list of summaries containing the previously trained summaries and the expert summaries $(\hat{s}(x), \hat{s}_1(x), \dots, \hat{s}_n(x))$.

\subsection{Constructing Estimators}

Next, we use the generated summaries to construct estimators of the underlying parameters (corresponding to step \textcircled{\scriptsize{2}} in Figure~\ref{fig:overview}). To do so, we first perform the following series expansion:
\begin{equation}
\hat{\theta}_i(x) = a_i + b_{ij}\hat{s}^j(x) + c_{ijk}\hat{s}^j(x)\hat{s}^k(x) + \dots, \label{eq:expansion}
\end{equation}
where we used the Einstein summation convention, meaning that we sum over all indices that appear as lower and upper index in a term. One can choose the coefficients $a_i$, $b_{ij}$, etc. such that they minimize the bias of the resulting estimator around the chosen fiducial point $\theta_\mathrm{fid}$. In this work, we will only consider first-order expansions, however, the construction of higher-order terms follows the same principle. We start by expanding the expectation value of the summaries around the fiducial parameter
\begin{align*}
\Psi_\theta(\hat{s}) = \Psi_{\theta_\mathrm{fid}}(\hat{s}) + \left.\frac{\partial\Psi_\theta(\hat{s})}{\partial\theta}\right\vert_{\theta=\theta_\mathrm{fid}}(\theta - \theta_\mathrm{fid}) + \mathcal{O}(\Delta\theta^2),
\end{align*}
where we defined $\Delta\theta = \theta - \theta_\mathrm{fid}$. Plugging this into the first-order expansion of equation \eqref{eq:expansion}, one can find the coefficients 
\begin{align*}
a_i &= \theta_{i,\mathrm{fid}} - \left(\frac{\partial\Psi_{\theta_\mathrm{fid}}(\hat{s})}{\partial\theta}\right)^{-1}_{ij}\Psi^j_{\theta_\mathrm{fid}}(\hat{s}) \\ b_{ij} &= \left(\frac{\partial\Psi_{\theta_\mathrm{fid}}(\hat{s})}{\partial\theta}\right)_{ij}^{-1}
\end{align*}

such that
\begin{equation*}
\E_{p(x\vert\theta)}[\hat{\theta}(x)] = \theta + \mathcal{O}(\Delta\theta^2 ).
\end{equation*}
In practice one has to estimate the coefficient $a_i$ and $b_{ij}$ from the sampled simulations $x$ of the fiducial parameter, leading to an additional small error from this approximation. We briefly note here, that the construction of higher-order estimators does not only require the estimation of the expected values $\Psi_{\theta_\mathrm{fid}}(\hat{s})$ and its derivative, but also from higher-order moments. We show how to construct a second order estimator, having an error of order $\mathcal{O}(\Delta\theta^3)$, in appendix \ref{ap:second_order}.

Since the construction of the estimators requires only simulations of the fiducial parameter and its perturbations, the simulations from the other parameters can be used to measure the performance of the constructed estimators.

Finally, although the approach we described uses a single estimation point, it can simply be extended to combine several estimation points. If one has access to multiple fiducial points (and perturbations of them), one can construct approximately unbiased estimators for each point. The different estimators can either come from the same underlying network or one can train a different network for every fiducial point individually. The constructed estimators can then be combined such that the bias and variance of the final estimators is minimized. For instance, \cite{estcomb2014} proposed to use a linear combination.

\section{Parameter Inference}
\label{sec:inference}

In this section, we show how one can use the previously described estimators to constrain the underlying model parameters using parameter inference (second part of the pipeline in Figure~\ref{fig:overview}). The goal of parameter inference is to find confidence regions of the underlying model parameter $\theta$, given an observation $x_\mathrm{obs}$ (corresponding to step \textcircled{\scriptsize{3}} in Figure~\ref{fig:overview}). This is no trivial task in practice, even if the posterior distribution $p(\theta\vert x_\mathrm{obs})$ is known, because it requires its integration to derive marginal distributions.

\subsection{Common Approaches}

\subsubsection{Markov Chain Monte Carlo} 
A popular way of performing parameter inference with a known likelihood function is to use Markov chain Monte Carlo (MCMC). However, even if the posterior distribution $p(\theta\vert x_\mathrm{obs})$ is known, running a MCMC can be unfeasible if the evaluation of the likelihood is expensive. A possible solution to this issue was proposed by \cite{Pelle2020}. They used Gaussian process regression to emulate the expensive posterior $p(\theta\vert x_\mathrm{obs})$ and performed a MCMC on the emulated posterior to obtain the parameter constraints. The obtained one and two dimensional marginal distributions from the Gaussian Process emulation were in excellent agreement with the ground truth. However, this approach is not applicable if the analytic form of the likelihood, or good approximations of it, are not available.

\subsubsection{Neural Density Estimators} 
In the case of an unknown or intractable posterior distribution $p(\theta\vert x_\mathrm{obs})$ one has to use likelihood free inference (LFI). A promising LFI technique is based on neural density estimators, as for instance proposed in \cite{Alsing2019}, where a combination of mixture density models and auto-regressive flows is used to learn the distribution of the summaries $p(\hat{s} \vert \theta)$. The posterior distribution $p(\theta\vert s_\mathrm{obs})$ can then be obtained by using Bayes theorem. We use the implementation of this method, called \texttt{PyDelfi}\footnote{\url{https://github.com/justinalsing/pydelfi}}, as baseline for our experiments in section \ref{sec:experiments}. The advantage of \texttt{PyDelfi} is that it can be applied to our settings as it does not require any on-the-fly simulations. Further, the constructed estimators, described in section \ref{sec:estimation}, are usually close to the actual parameter making it much simpler to learn the distribution $p(\hat{\theta}(x) \vert \theta)$, as opposed to summaries whose distribution is completely unknown or have a higher dimensionality.

\subsubsection{Approximate Bayesian Computation} 
Another common way of performing LFI is called approximate Bayesian computation (ABC). In ABC one usually replaces the high dimensional data $x$ with a low dimensional summary $\hat{s}(x)$. The ABC approximation of the posterior (e.g. see \cite{Fan2018}) is then given by
\begin{equation}
p_{ABC}(\theta\vert s_\mathrm{obs}) \propto \int K_h(\Vert s - s_\mathrm{obs}\Vert) p(s\vert\theta)p(\theta)\mathrm{d}s, \label{eq:ABC_post}
\end{equation}
where $s_\mathrm{obs} = \hat{s}(x_\mathrm{obs})$ is the summary of the observation, $K_h(x) = K(x/h)/h$ is a kernel density function with scale parameter $h$, $\Vert\cdot\Vert$ defines a norm and $p(\theta)$ denotes the prior distribution of the model parameters. It is straightforward to see that the ABC posterior converges to $p(\theta\vert s_\mathrm{obs})$ for $h\to 0$. If the summary $\hat{s}(x)$ is sufficient, we even have
\begin{equation}
\lim_{h\to 0}p_{ABC}(\theta\vert s_\mathrm{obs}) = p(\theta\vert s_\mathrm{obs}) = p(\theta\vert x_\mathrm{obs}).
\end{equation}
The simplest ABC algorithm is called rejection ABC which works similarly to MCMC. It first generates a random sample $\theta$ from $p(\theta)$ and then generates a sample $x$ from $p(x\vert\theta)$. The new parameter $\theta$ is accepted if $\Vert\hat{s}(x) - s_\mathrm{obs}\Vert < \epsilon$, where the threshold $\epsilon$ is a hyper-parameter corresponding to the scale $h$. Using this technique, one can generate samples from the ABC posterior and use them to calculate the marginal distributions and eventually the confidence regions. The downside of this algorithm is that it requires a large number of simulations $x$, even with a carefully selected threshold. More advanced ABC algorithms such as population Monte Carlo ABC (PMC-ABC) \citep{PMCABC} require fewer on-the-fly simulations, but can not be applied to a fixed grid of simulations. 

\subsection{Gaussian Process ABC}

A combination of Gaussian process regression and ABC has been proposed by \cite{Jarven2017}, using the Gaussian process to emulate the discrepancy measure of the ABC. Later, \cite{Jarven2019} used Gaussian process regression to emulate (noisy) likelihood estimates. However, they mainly consider a synthetic likelihood setting where the unknown likelihood is approximated by a Gaussian distribution, focusing on the error estimation of the posterior and acquisition strategies for new simulations. In this work we compute ABC posterior estimates and their corresponding uncertainties using the estimator presented in Section~\ref{sec:estimation}. Afterwards, we use Gaussian process regression to emulate the posterior similarly to \cite{Jarven2019}. This does not require on-the-fly simulations as in standard ABC methods but still allows likelihood free inference. Further, it makes it possible to directly learn the posterior $p(\theta\vert \hat{\theta}_\mathrm{obs})$ conditioned on a set observation $\hat{\theta}_\mathrm{obs}$ and does not require to learn the conditional distribution $p(\theta\vert \hat{\theta})$ jointly in $\theta$ and $\hat{\theta}$ like neural density estimators. 

We use our available $n_x$ draws from the distribution $p(x\vert\theta)$ and utilize the constructed estimator of the previous section. We can then obtain an unbiased, consistent and non-negative Monte Carlo estimate of the ABC posterior described by equation \eqref{eq:ABC_post} (corresponding to step \textcircled{\scriptsize{4}} in Figure~\ref{fig:overview}) via
\begin{equation}
\hat{p}_{ABC}(\theta\vert \hat{\theta}_\mathrm{obs}) \propto \frac{p(\theta)}{n_x}\sum_{\alpha=1}^{n_x}K_h(\Vert \hat{\theta}_\alpha - \hat{\theta}_\mathrm{obs}\Vert), \label{eq:ABC_est}
\end{equation}
where we used our estimates $\hat{\theta}_\alpha \sim p(\hat\theta\vert\theta)$ as summaries and the normalization is irrelevant for our purpose. Note that the scaling parameter $h$ can be used to smooth the ABC posterior if the fixed grid is too coarse. A good choice for the norm inside the kernel density function is the Fisher distance \citep{Charnock2018} around the fidicual parameter
\begin{equation*}
\Vert \hat{\theta}_\alpha - \hat{\theta}_\mathrm{obs}\Vert^2 = (\hat{\theta}_\alpha - \hat{\theta}_\mathrm{obs})^T \hat{\Im}_\mathrm{fid} (\hat{\theta}_\alpha - \hat{\theta}_\mathrm{obs}),
\end{equation*} 
where we defined the estimated Fisher matrix
\begin{equation*}
    \hat{\Im}_\mathrm{fid} = \left.\frac{\partial\Psi_\theta(\hat{s})}{\partial\theta}\right\vert_{\theta=\theta_\mathrm{fid}}^T\mathrm{Cov}_{\theta_\mathrm{fid}}(\hat{s})^{-1}\left.\frac{\partial\Psi_\theta(\hat{s})}{\partial\theta}\right\vert_{\theta=\theta_\mathrm{fid}}.
\end{equation*}
For a first order estimator this is equivalent to the Mahalanobis distance of the summary and scales the entries of the summary according to their variances. As \cite{Pelle2020} pointed out, performing Gaussian Process regression on the posterior $p(\theta\vert \hat{\theta}_\mathrm{obs})$ itself is difficult. One reason for this, is that the posterior is usually close to zero in a large subset of the parameter space $\Theta$. This can be alleviated by performing the regression on the log-posterior $\log(p(\theta\vert \hat{\theta}_\mathrm{obs}))$ instead. Using equation \eqref{eq:ABC_est} we can construct an estimator of the ABC log-posterior 
\begin{equation}
\hat{\mathcal{L}}_{ABC}(\theta) = \log\left(\frac{p(\theta)}{n_x}\sum_{\alpha=1}^{n_x}K_h(\Vert \hat{\theta}_\alpha - \hat{\theta}_\mathrm{obs}\Vert\right),
\end{equation}
where we left out a constant offset coming from the normalization constant of the ABC posterior estimate. One should note that this estimator is still consistent, but not unbiased. The bias follows directly from Jensen's inequality and becomes negligible for sufficiently large sample sizes $n_x$. Another issue with normal Gaussian process regression is that the variance of the estimators depends on the underlying parameter $\theta$ and is not constant over the parameter space $\Theta$. Using \texttt{gpflow}\footnote{\url{https://www.gpflow.org/}} \cite{Mathews2017} we adopt the following model for the Gaussian process regression (corresponding to step \textcircled{\scriptsize{5}} in Figure~\ref{fig:overview})
\begin{align*}
\mathcal{L}_{ABC} &\sim \mathcal{GP}(0,k(\cdot ,\cdot )) \\
\hat{\mathcal{L}}_{ABC}(\theta_\alpha) \vert \mathcal{L}_{ABC}, \theta_\alpha &\sim \mathcal{N}\left(\mathcal{L}_{ABC}(\theta_\alpha), \sigma_\alpha^2\right),
\end{align*}
where $k(\cdot ,\cdot )$ is a kernel function and $\sigma_\alpha^2$ is the estimated variance of the ABC log-posterior estimate. To obtain the variances $\sigma_\alpha^2$ we use a Taylor series expansion of the moments. If $g:\mathcal{U}\subseteq\mathbb{R}\rightarrow\mathbb{R}$ is a smooth function and $X\in\mathcal{U}$ a continuous random variable with finite mean and variance $\mu_X, \sigma^2_X < \infty$, one can approximate the variance of the transformed random variable as
\begin{equation}
\mathrm{Var}\left[g(X)\right] \approx \left.\frac{\mathrm{d}g(x)}{\mathrm{d}x}^2\right\vert_{x=\mu_X}\sigma_X^2.
\end{equation}
We find this to be a good approximation\footnote{Using the mentioned bound, the next order term is approximately 10 times smaller for a Gaussian distributed $X$.} for $g(x) = \log(x)$ if $X$ is approximately normal distributed and $\mu_X / \sigma_X \gtrsim 2.5$. Therefore, we start by estimating the variance of our ABC posterior estimates by taking the variance on both side of equation \eqref{eq:ABC_est}
\begin{equation*}
\mathrm{Var}\left[\hat{p}_{ABC}(\theta_\alpha\vert \hat{\theta}_\mathrm{obs})\right] \approx \frac{p(\theta_\alpha)^2}{n_x}\mathrm{Var}\left[K_h(\Vert \hat{\theta}_\alpha - \hat{\theta}_\mathrm{obs}\Vert\right],
\end{equation*}
and then propagate this variance to the log-posterior via the Taylor series expansion
\begin{equation}
\sigma_\alpha^2 = \frac{p(\theta_i)^2}{n_x}\frac{\mathrm{Var}\left[K_h(\Vert \hat{\theta}_\alpha - \hat{\theta}_\mathrm{obs}\Vert\right]}{\hat{p}_{ABC}(\theta\vert \hat{\theta}_\mathrm{obs})^2}.
\end{equation}
The evaluation of the Gaussian process with optimized kernel parameters is computationally cheap, making it straight forward to obtain all relevant properties of the posterior distribution $p_{ABC}(\theta\vert \hat{\theta}_\mathrm{obs})$ with standard methods like MCMC.

\section{Experiments}
\label{sec:experiments}

To test our method, we apply it to the inference of cosmological parameters from  weak gravitational lensing maps. For this purpose, we consider simplified Gaussian maps for which the sufficient summary statistics is known. In weak lensing, one measures the shapes of millions of galaxies across the sky that have been distorted by the density inhomogeneities in the universe on the line-of-sight, according to general relativity. The correlations between the measured shapes can then be used to construct convergence maps, which are maps of the mass overdensity projected on the sky. Convergence maps can be generated using large, computationally expensive, cosmological simulations. These simulations are usually so expensive to run that it is impossible to run them on-the-fly during the data analysis. Comprehensive reviews about weak gravitational lensing can be found in \cite{WLreview,wlrevie99}. 

Common ways to analyze weak lensing convergence maps usually rely on two-point correlation functions (e.g. \cite{kids1000_res, desy3_res}). However, it is known that the two-point correlations function is not a sufficient summary of the data and there is an on going search for so called non-Gaussian statistics (e.g. \cite{Dominik2020}). This problem is going to become more significant as the quality and quantity of the observed data is going to increase in future surveys like Eulcid\footnote{\url{https://www.euclid-ec.org/?page_id=79}}~\cite{euclid}. Replacing the human made summaries with automatically learned summaries that can be used for LFI can therefore be extremely beneficent to achieve tight and robust cosmological parameter constraints.

\subsection{Model Parameter} 

Our experiments are based on the flat wCDM cosmological model that extends the standard $\Lambda$CDM model with the dark energy equation of state parameter $w_0$. The remaining parameters of the $\Lambda$CDM are the total matter density fraction $\Omega_M$, the baryonic matter density fraction $\Omega_b$, the Hubble parameter $H_0 \equiv 100h$, the spectral index $n_s$ and the fluctuation amplitude $\sigma_8$. Here, flat is referring to the total energy density parameter being equal to one 
$\Omega = \Omega_\mathrm{CDM} + \Omega_b + \Omega_\Lambda \equiv 1$, where $\Omega_\mathrm{CDM} = \Omega_M - \Omega_b$ is the dark matter density fraction and the dark energy density fraction $\Omega_\Lambda$ is determined by the constraint $\Omega = 1$, which corresponds to a geometrically flat universe. We consider two cases, i) a two-dimensional model (hereafter \textbf{2D}) where we only vary the parameters weak lensing is most sensitive to, namely $\Omega_M$ and $\sigma_8$; and ii) a six-dimensional model (hereafter \textbf{6D}) that allows all six parameters to vary. 

\begin{table}[b]
\caption{Overview of the chosen fiducial parameters $\theta_{\mathrm{fid},i}$, their perturbations $\Delta\theta_i$ and uniform priors. The degeneracy parameter $S_8 = \sigma_8\sqrt{\Omega/0.3}$ is a derived parameter and captures the correlations between $\Omega_M$ and $\sigma_8$.  \label{tb:params}}
\begin{center}
\begin{tabular}{lrrl}
$\theta_i$ & $\theta_{\mathrm{fid},i}$ & $\Delta\theta_i$ & Prior \\
\hline \vspace{-5pt}\\ 
$\Omega_M$ & 0.3175 & 0.01 & $[0.25, 0.4]$ \\
$\Omega_b$ & 0.049 & 0.002 & $[0.03, 0.07]$ \\
$h$ & 0.6711 & 0.02 & $[0.5, 0.9]$ \\
$n_s$ & 0.9624 & 0.02 & $[0.8, 1.2]$ \\
$\sigma_8$ & 0.834 & 0.015 & $[0.72, 0.92]$ \\
$w_0$ & -1.0 & 0.05 & $[-1.3, -0.7]$ \\
$S_8$ & 0.858 & N/A & $[0.79, 0.94]$ \\
\end{tabular}
\end{center}
\end{table}

We chose the same fiducial parameters and perturbations as \cite{Quijote} for the training, which are listed in table \ref{tb:params}. Note that we place an additional prior on the degeneracy parameter $S_8 = \sigma_8\sqrt{\Omega/0.3}$, corresponding to the width of the degenerate contours (as shown in Figure~\ref{fig:cons}).

\subsection{Mock-Survey Settings} 

\begin{figure}[hbt]
\centering
\includegraphics[width=1.\linewidth]{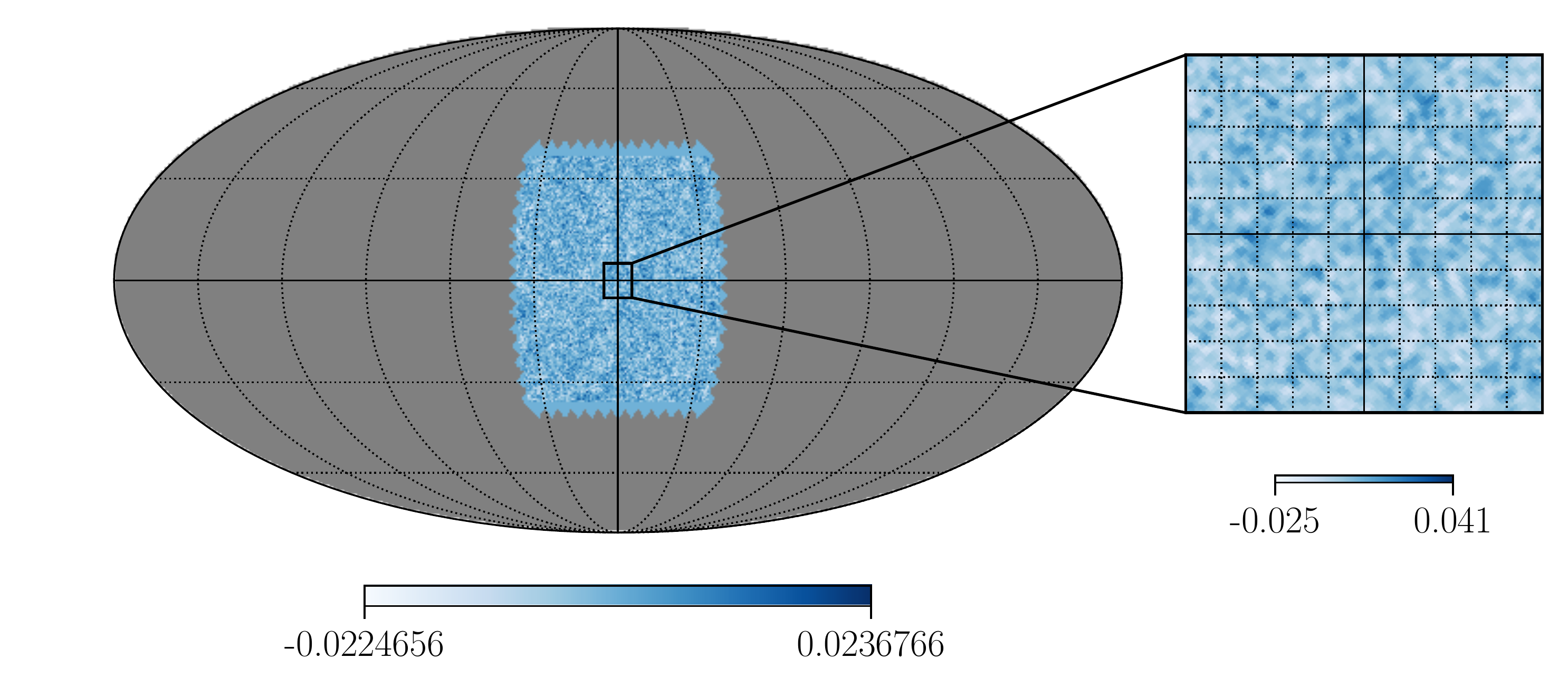}
\caption{A GRF mimicking a convergence map of our mock survey generated for our fiducial parameter. \label{fig:zoom}}
\end{figure}

We consider a stage-III like survey, meaning that the convergence maps span $\sim5000\deg^2$ on the sky (see Figure~\ref{fig:zoom}). Further, instead of using convergence maps generated from cosmological simulations, we use Gaussian random fields (GRF). The advantage of GRF  is, that while being similar to realistic convergence maps, they are easy to model statistically. A spherical GRF is completely determined by its spherical harmonic power spectrum (see appendix \ref{ap:GRF} for more details), which means that the power spectrum is a sufficient statistic and that one can use power spectra to generate GRF. The power spectrum of convergence maps is itself a summary statistic often used for cosmological parameter inference\cite{kids1000_res,desy3_res} and advanced codes exist that allow the fast generation of spectra ($\sim 1$ second) for any set of cosmological parameters. In our  model we use the publicly available code \texttt{CCL}\footnote{\url{https://github.com/LSSTDESC/CCL}} (The Core Cosmology Library) described in \cite{CCL} to predict the spectra. 

\subsubsection{Redshift Distribution}

\begin{figure}
    \centering
    \includegraphics[width=1.0\linewidth]{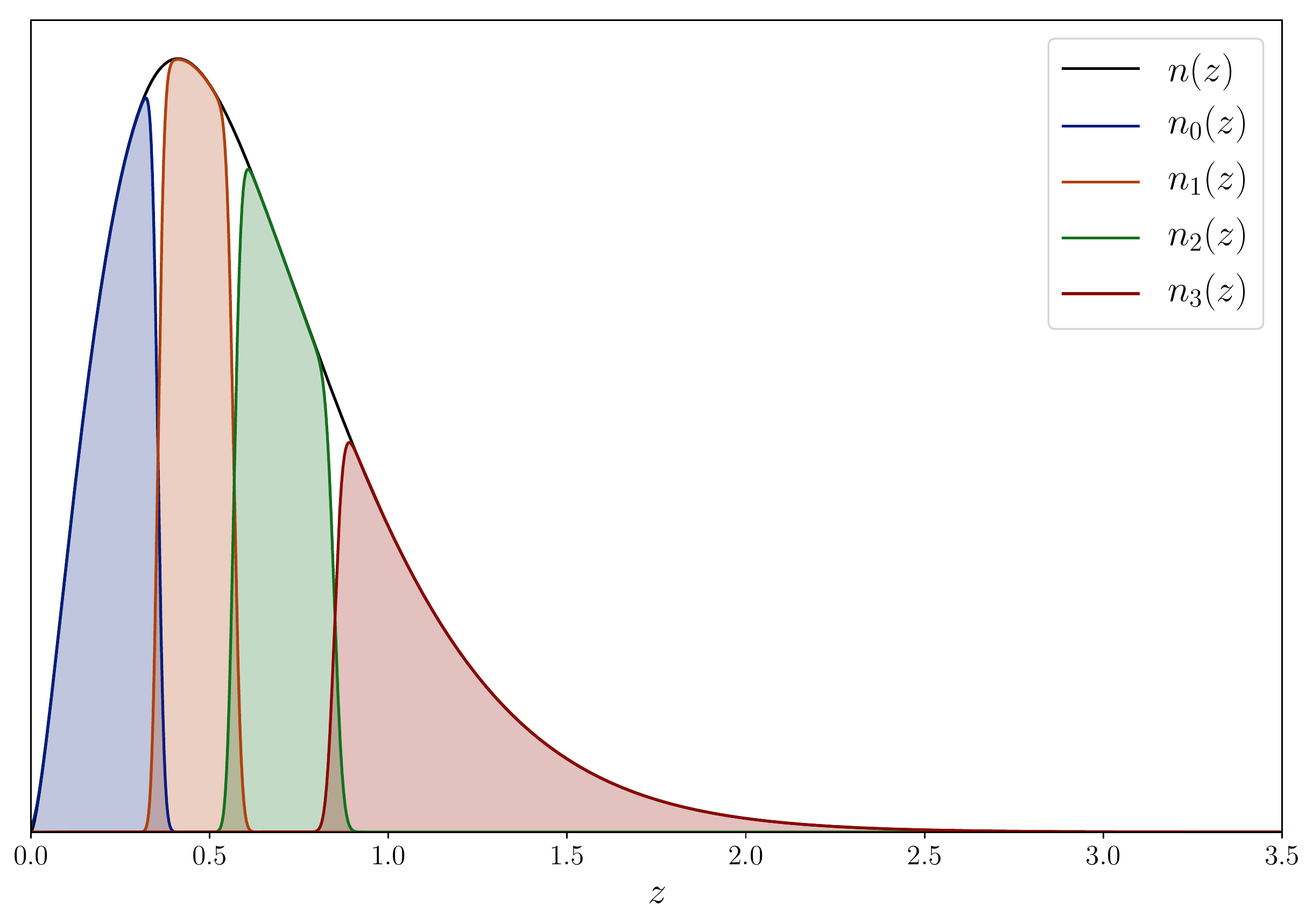}
    \caption{Redshift distributions of the individual bins and the total redshift distribution of the considered mock survey. \label{fig:red_dist}}
\end{figure}

The generation of convergence power spectra does not only require the cosmological parameters, but also observational parameters. The most important observational parameter is the galaxy distribution along the line-of-sight of the observer, i.e. how likely it is to observe a galaxy at a certain distance (redshift) from the observer, called the redshift distribution. We use the same total redshift distribution as \cite{Dominik2020}, which has an analytic formula
\begin{equation*}
    n(z) \propto z^{1.5}\exp\left(-\left(\frac{z}{0.31}\right)^{1.1}\right),
\end{equation*}
and is a realistic redshift distribution for a stage-III cosmological survey. The normalization of the distribution depends on the redshift boundaries of the survey (maximum observed distance). For our model we chose $z_\mathrm{min} = 0.0$ (position of the observer) and $z_\mathrm{max} = 3.5$. Redshift distributions are normalized like any probability distribution by the condition
\begin{equation*}
    \int_{z_\mathrm{min}}^{z_\mathrm{max}} n(z) \mathrm{d}z = 1.
\end{equation*}
Further, we consider a survey setting where the observed galaxies are assigned into four tomographic bins according to their redshift. This allows the creation of four convergence maps, all having a different redshift distribution. Note however, that their features are correlated in a certain way, since they stem from the same universe. These correlation are defined by their cross-spectra, which can also be predicted using \texttt{CCL}.
Each of the tomographic bins should have approximately the same number of galaxies. Optimally, one could measure the redshift of a galaxy exactly, such that redshift distributions of the individual bins follow the same redshift distribution as the survey truncated with different $z_\mathrm{min}$ and $z_\mathrm{max}$. However, in reality redshift measurements are prone to errors such that the individual redshift distributions broaden. We model this measurement error following \cite{Amara2007} assuming that the standard error of a redshift measurement is given by $\sigma_z = \delta(1 + z)$, with $\delta = 0.01$. The resulting redshift distributions of the individual bins are shown in Figure~\ref{fig:red_dist} and do not have a closed form expression.

\subsubsection{Mock-Data Generation}

Generating a GRF with a single power spectrum is fairly straight forward. However, generating different GRF with given cross spectra requires a more advanced generation procedure. The necessary algorithm is described in appendix C of \cite{Sgier2020}. To actually generate the spherical GRF we use the Hierarchical Equal Area iso-Latitude Pixelization tool (\texttt{HEALPix}\footnote{\url{http://healpix.sourceforge.net}}) described in \cite{Healpix}, using \texttt{CCL} to calculate the input power- and cross-spectra of convergence maps given a set of cosmological parameters and the necessary redshift distributions. Our cosmological model used \texttt{CCL} to predict spectra up to $\ell_{\max}  = 1000$. In \texttt{Healpix} the resolution of a map is given by the $\mathtt{nside}$ parameter, which we fixed for all experiments to $\mathtt{nside} = 512$, leading to $\sim 400,000$ relevant pixel (see Figure~\ref{fig:zoom}) for each of the four redshift bins.

 We generated $10,000$ simulations with the fiducial parameter and each of 12 individual perturbations. Additionally, we generated $n_x = 50$ simulations for $n_\theta = 1024$ different parameters sampled as a Sobol sequence inside the reported priors for the \textbf{2D} model and $n_\theta = 2048$ for the \textbf{6D} model.

\subsubsection{Observational Noise}
\label{sec:obs_noise}

The measured data in realistic cosmological surveys contains a substantial amount of noise. The observational noise can be modeled as Gaussian random noise \citep{noiseref} that mainly depends on the number of observed galaxies. The distribution is centered at zero and its standard deviation is given by
\begin{equation*}
\sigma_\mathrm{noise} = \frac{\sigma_e}{\sqrt{An}},
\end{equation*}
where $\sigma_e = 0.3$ is a typically measured ellipticity dispersion, $A$ is the area of a given pixel and $n$ is the galaxy number density. We add observational noise to all our simulations with a chosen galaxy density of $n = 12$ galaxies/arcmin${^2}$ which translates to 3 galaxies/arcmin${^2}$ per redshift bin. Further, it can be shown that the power spectrum of a pure (pixelized) noise map is constant
\begin{equation}
    \mathcal{C}_\mathrm{noise} = \frac{f_\mathrm{sky}}{n_\mathrm{pix}}4\pi \sigma_\mathrm{noise},
\end{equation}
where $f_\mathrm{sky}$ corresponds to the fraction of the sky that is covered by the survey and $n_\mathrm{pix}$ to the number of pixels required to cover the whole sphere. 

\subsection{GCNN Network} 
\label{sec:GCNN}

\begin{figure}[b!]
    \centering
    \includegraphics[width=0.5\textwidth]{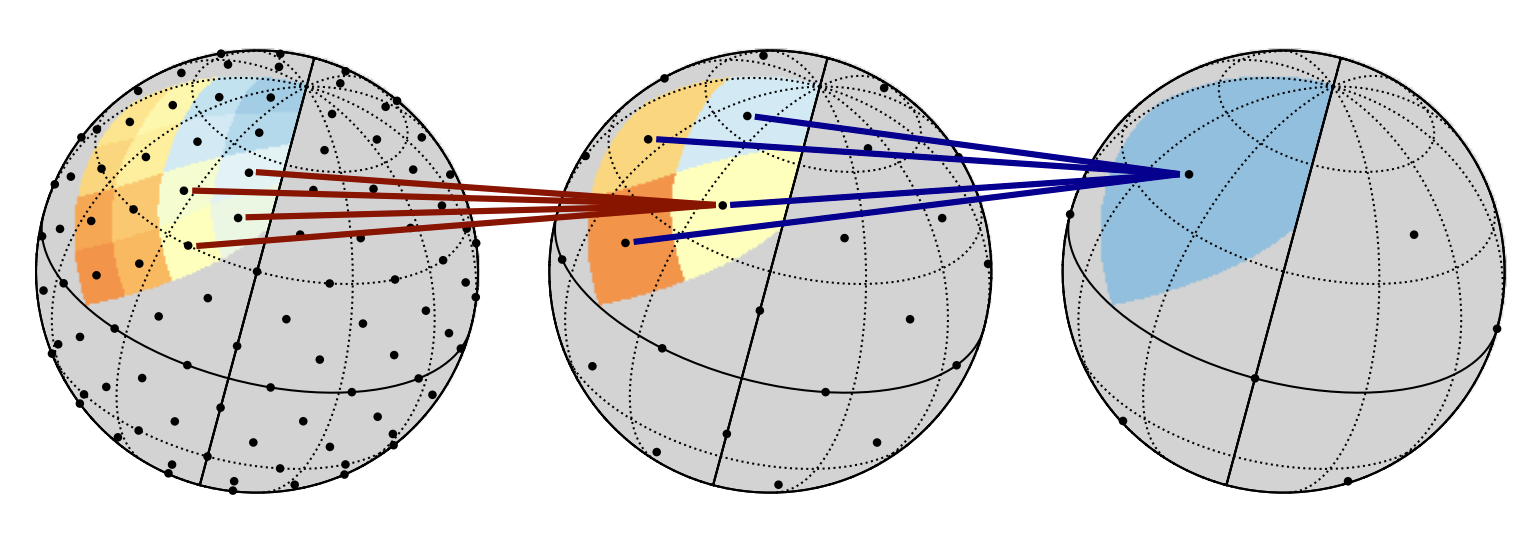}
    \caption{Downsampling procedure using the hierarchical pixelization approach of \texttt{HealPix}. Four pixels are combined into one. The weights of the combination can be learned during the training of the network. Figure~taken from \cite{deepsphere_cosmo}.  \label{fig:healpix}}
\end{figure}

The chosen survey configuration of the model makes it impossible to treat the data as flat images. There exist multiple approaches to deal with spherical data. We chose to use \texttt{DeepSphere}\footnote{\url{https://github.com/deepsphere/deepsphere-cosmo-tf2}}, described in \cite{deepsphere_cosmo}, that treats the pixels of the the generated GRF as nodes in a graph. The distances between the nodes are given by the spherical geometry of the maps. The graph convolutional filters in \texttt{DeepSphere} are expressed in terms of Chebyshev polynomials and applied to the data using the graph Laplacian. Additionally it uses  the hierarchical pixelization approach of \texttt{HealPix} to downsample the data while maintaing is spherical geometry. This downsampling procedure is refered to as pseudo convolutions and displayed in Figure~\ref{fig:healpix}. However, to use this feature one has to add a padding to the input of the GCNN such that a valid downsampling is possible (see edges of the patch shown in Figure~\ref{fig:zoom}).

\begin{figure}[ht]
    \centering
    \includegraphics[width=0.2\textwidth]{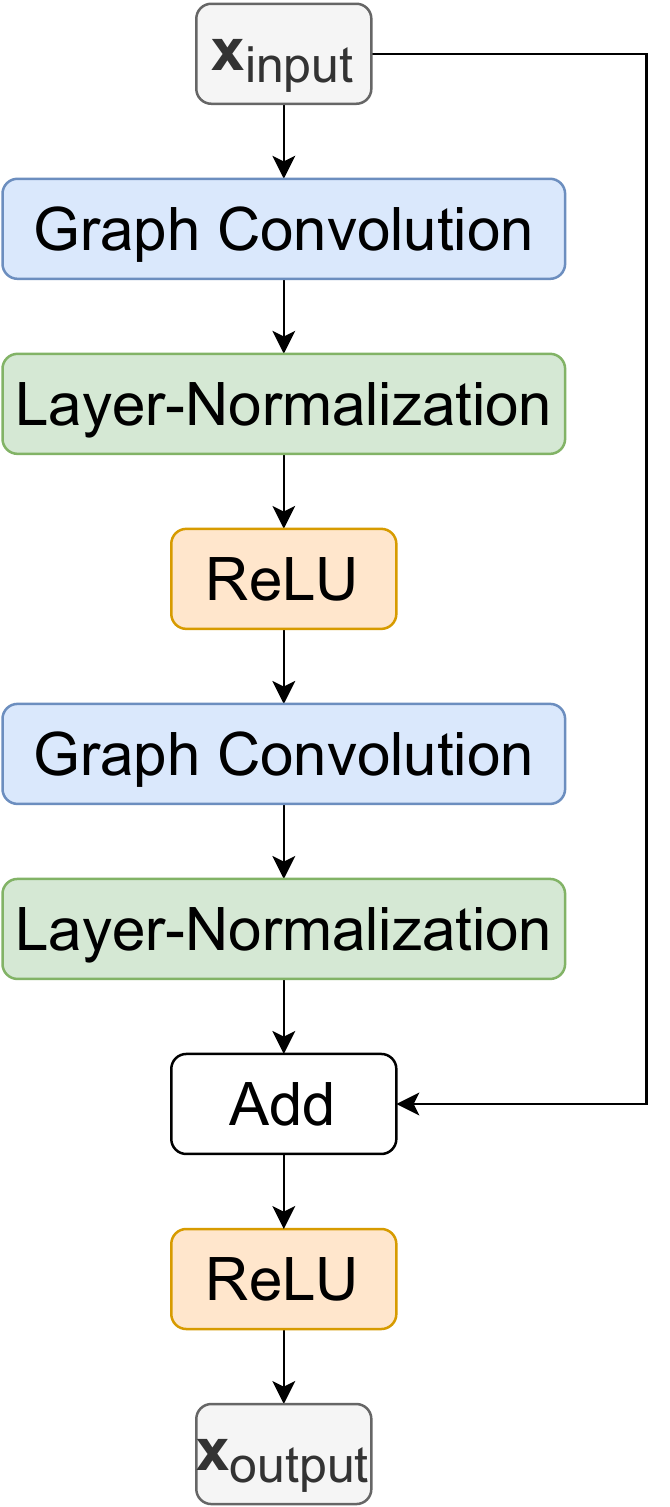}
    \caption{Architecture of the used residual layer.  \label{fig:reslayer}}
\end{figure}

The architecture of the used network is shown in table \ref{tb:architecture}. All presented layer use the ReLU activation function. The architecture of the residual layers is shown as a flowchart in Figure~\ref{fig:reslayer}. All graph convolutions use Chebyshev polynomials with degree $K=5$. Further details and a concrete implementation can be found on our \texttt{github}\footnote{\url{https://github.com/jafluri/cosmo_estimators}} page.

\begin{table}[t]
    \centering
    \caption{Architecture of the used GCNN. We report layer type, output shape ($N_b$ being the batch size) and number of trainable parameters. The residual layer (see Figure~\ref{fig:reslayer}) is repeated ten times. The output shape depends on the model. For the \textbf{2D} model we have $N_\mathrm{out} = 2$ and $N_\mathrm{out} = 6$ for the \textbf{6D} model. \label{tb:architecture}}
    \begin{tabular}{llr}
         \textbf{Layer Type} & \textbf{Output Shape} & \textbf{\# Parameter} \\
         \hline\vspace{-5pt} \\ 
         Input & ($N_b$,\ 428032,\ 4) & 0 \\
         Pseudo Convolution & ($N_b$, 107008, 16) & 272 \\
         Pseudo Convolution & ($N_b$, 26752, 32) & 2080 \\
         Pseudo Convolution & ($N_b$, 6688, 64) & 8256 \\
         Pseudo Convolution & ($N_b$, 1672, 128) & 32896 \\
         Layer-Normalization & ($N_b$, 1672, 128) & 256 \\
         Residual Layer & ($N_b$, 1672, 128) & 170528 \\
         \multicolumn{3}{c}{$\vdots$} \\
         Residual Layer & ($N_b$, 1672, 128) & 170528 \\
         Pseudo Convolution & ($N_b$, 418, 128) & 32896 \\
         Flatten & ($N_b$, 53504) & 0 \\
         Layer-Normalization & ($N_b$, 53504) & 107008 \\
         Fully Connected &  ($N_b$, $N_\mathrm{out}$) & 53504$N_\mathrm{out}$
    \end{tabular}
\end{table}

\begin{figure*}
\includegraphics[width=1.0\linewidth]{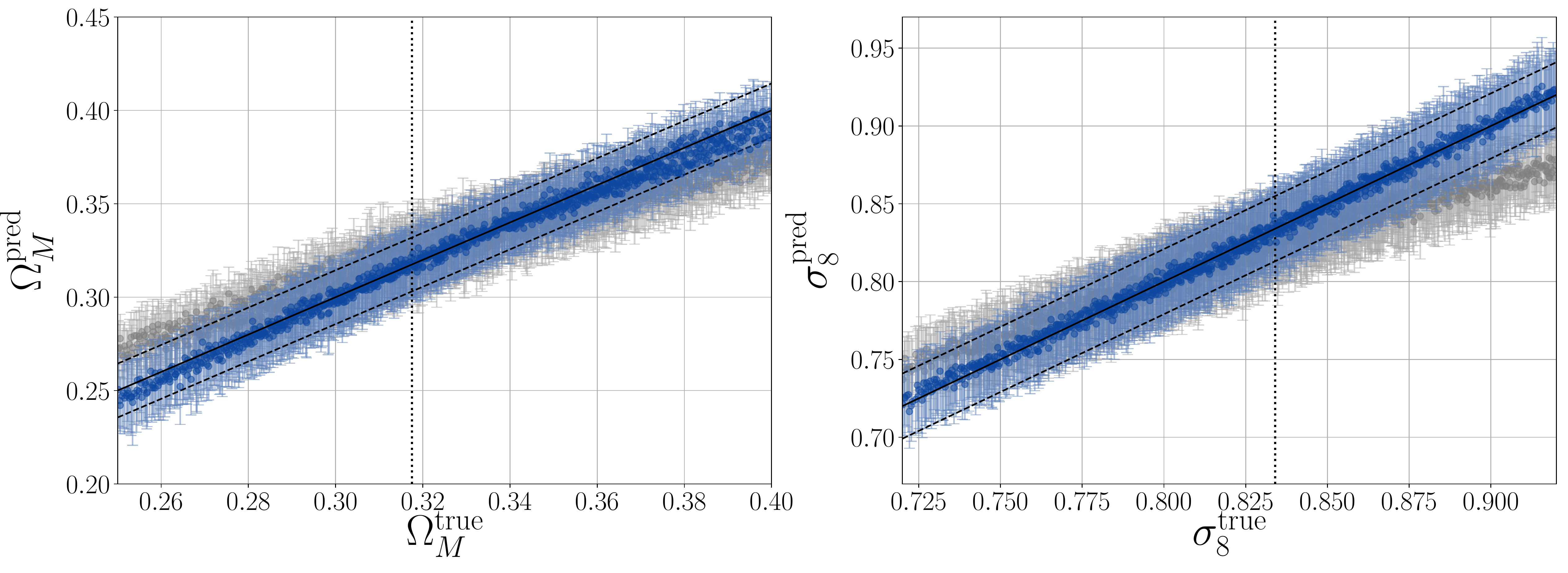}
\caption{Mean predictions and their standard deviation of the test set of the \textbf{2D} model in blue and the model trained with the MSE loss (equation \eqref{eq:square_loss}) in grey. The vertical dotted line indicates the fiducial paramter used for the training of the \textbf{2D} model. The solid line indicates an unbiased prediction $y = x$, and the dashed lines above and below indicate the lower bound on the standard deviation for any estimator of the fiducial parameter. \label{fig:true_vs_pred}}
\end{figure*}

\subsubsection{Training}
\label{sec:GCNN_train}

All graph convolutional neural networks presented in this work were trained for 100,000 steps with batch size of $128$ distributed among 32 GPUs on the cluster computer Piz Daint\footnote{\url{https://www.cscs.ch/computers/piz-daint/}}. This batch size has to be multiplied with $2d_\theta + 1$, because of the necessary perturbations to estimate the derivatives with respect to the underlying parameters. Beside the loss defined in equation \eqref{eq:loss_1} we added the regularization term of equation \eqref{eq:regu_loss_3} with parameter $\lambda = 10$. The observational noise (see section \ref{sec:obs_noise}) modeled as Gaussian random noise was sampled anew for each training step, but the same seed was used for each perturbation. The weights were optimized using the Adam optimizer \citep{Adamopt} with initial learning rate of 0.0001 and moments $\beta_1=0.9$ and $\beta_2=0.999$. Additionally, to avoid large weight updates, we applied global gradient clipping, reducing the global norm of the gradients in a training step to 5.0 if necessary.

\subsection{Ground Truth} 

The advantage of this model is that a sufficient summary statistics is known from the beginning. All maps are generated from the power spectra  calculated with \texttt{CCL}. Let $\mathcal{C}(\theta)$ be such a spectrum, already corrected for the survey geometry (see appendix \ref{ap:suvey_geo}). Given a spectrum $\mathcal{C}_\mathrm{obs}$ measured from an observation $x_\mathrm{obs}$, the posterior distribution is given by \citep{selletin2016}
\begin{equation}
\label{eq:true_post}
p(\theta \vert \mathcal{C}(\theta), \mathcal{C}_\mathrm{obs}, S) \propto \frac{\det(S)^{-\frac{1}{2}}}{\left[1 + \frac{Q}{N-1}\right]^{\frac{N}{2}}},
\end{equation}
where $Q$ is defined as 
\begin{equation*}
Q = (\mathcal{C}(\theta) + \mathcal{C}_\mathrm{noise} - \mathcal{C}_\mathrm{obs})^TS^{-1}(\mathcal{C}(\theta) + \mathcal{C}_\mathrm{noise} - \mathcal{C}_\mathrm{obs}),
\end{equation*}
the covariance matrix $S$ is estimated from the $N = 10,000$ simulations of the fiducial parameter and $\mathcal{C}_\mathrm{noise}$ is the noise contribution, which does not depend on the cosmological parameters. We make the common assumption that the covariance matrix $S$ does not depend on the cosmological parameters $\theta$. For parameter inference, one can directly integrate the posterior distribution with a MCMC. For parameter estimation one can use the likelihood to calculate the Fisher information matrix, which in turn gives the smallest possible variance of any estimator. It can be shown \citep{Lange1989}, that the Fisher information matrix of the likelihood defined in equation \eqref{eq:true_post} is given by
\begin{equation}
\mathbf{I}_{ij}(\theta) = \frac{N}{N + 2}\frac{\partial \mathcal{C}(\theta)}{\partial \theta_i}^T S^{-1} \frac{\partial \mathcal{C}(\theta)}{\partial \theta_j}.
\end{equation}
Leading to a lower bound on the variance of any unbiased estimator $\mathrm{Var}\left[\hat{\theta}_i\right] \geq (\textbf{I}^{-1})_{ii}$. We calculate the lower bound for the fiducial parameters using the perturbed spectra to numerically approximate the derivatives.

\subsection{Parameter Estimation}

We trained a GCNN (see section \ref{sec:GCNN}) for $100,000$ training steps using a batch size of 128. Afterwards, we used the $10,000$ simulations of the fiducial parameter and its perturbations to construct a first order estimator. 

Additionally, inspired by \cite{sum1,sum2}, we trained the same network with the mean squared error loss (MSE) defined in equation~\eqref{eq:square_loss} for $100,000$ training steps with a batch size of 128. The predictions of this model where used as a benchmark for the parameter estimation task.

We compare the true parameters and the mean predictions of the constructed estimator of the \textbf{2D} model and the predictions of the MSE model for the $n_x = 50$ maps generated for the $n_\theta = 1024$ test parameters in Figure~\ref{fig:true_vs_pred}. Even though the \textbf{2D} model network was only trained on the fiducial parameter and its perturbations, the mean predictions are not significantly biased throughout the considered prior range. Further, the variance of the constructed estimator is close to the calculated lower bound of the known sufficent summary. A plot of the residues, i.e. the biases, can be found in appendix \ref{ap:biases}. The model trained with the MSE loss exhibits a larger bias despite being trained on the whole parameter space. This bias also increased towards the edges of our chosen priors as expected and shown in \cite{ribli2019weak} and \cite{Navaro2020}.

\subsection{Parameter Inference}

\begin{table*}[t]
\caption{Overview of the parameter inference results. The number of evaluations corresponds to the number of different parameter $\theta$ used to obtain the posterior distribution. The constraints report the mean values and the 68\% confidence regions. The JSD was calculated with respect to the MCMC run, which we consider as our ground-truth (lower values are better).  \label{tb:res}}
\begin{center}
\begin{tabular}{llccccccrr}
 &  & \multicolumn{6}{c}{\textbf{Constraints}} \\
\cline{3-8}
 & \textbf{Method} & $\Omega_M$ & $\Omega_b$ & $h$ & $n_s$ & $\sigma_8$ & $w_0$ & \textbf{JSD} & \textbf{Evaluations} \\
\hline \vspace{-5pt}\\ 
\multirow{4}{*}{\vspace*{-16pt}\rotatebox{90}{\textbf{2D}}} & MCMC      & $0.34_{-0.02}^{+0.01}$ & - & - & - & $0.80_{-0.02}^{+0.02}$ & - & 0.0 & $500,000$\vspace*{5pt}\\
 & GP ABC    & $0.35_{-0.02}^{+0.02}$ & - & - & - & $0.80_{-0.03}^{+0.03}$ & - & \bf{0.055} & 128 \vspace*{5pt}\\
 & PyDelfi   & $0.35_{-0.02}^{+0.02}$ & - & - & - & $0.80_{-0.03}^{+0.02}$ & - & 0.166 & 128 \vspace*{5pt}\\
 & MSE   & $0.35_{-0.03}^{+0.03}$ & - & - & - & $0.79_{-0.04}^{+0.04}$ & - & 0.323 & 128 \vspace*{5pt}\\
\hline \vspace{-5pt}\\ 
\multirow{3}{*}{\vspace*{-12pt}\rotatebox{90}{\textbf{6D}}} & MCMC      & $0.34_{-0.02}^{+0.02}$ &$0.05_{-0.01}^{+0.01}$ &$0.73_{-0.10}^{+0.10}$ &$0.94_{-0.06}^{+0.06}$ &$0.81_{-0.03}^{+0.03}$ &$-1.01_{-0.17}^{+0.17}$ & 0.0 & $10,000,000$\vspace*{5pt}\\
 & GP ABC    & $0.34_{-0.03}^{+0.03}$ &$0.05_{-0.01}^{+0.02}$ &$0.72_{-0.10}^{+0.13}$ &$0.93_{-0.08}^{+0.06}$ &$0.81_{-0.04}^{+0.03}$ &$-1.05_{-0.25}^{+0.09}$ & \bf{0.026}${}^\dagger$ & 2048 \vspace*{5pt}\\
 & PyDelfi   & $0.34_{-0.02}^{+0.02}$ &$0.05_{-0.02}^{+0.02}$ &$0.70_{-0.12}^{+0.11}$ &$0.94_{-0.07}^{+0.07}$ &$0.80_{-0.04}^{+0.03}$ &$-1.06_{-0.24}^{+0.08}$ & 0.028${}^\dagger$ & 2048 \vspace*{5pt}\\
\hline \vspace{-9pt}\\ 
 & \multicolumn{9}{l}{\footnotesize{${}^\dagger$ Mean Jensen-Shannon divergence of the 15 two-dimensional marginal distributions.}}
\end{tabular}
\end{center}
\end{table*}

\begin{figure}[t!]
    \centering
    \includegraphics[width=1.0\linewidth]{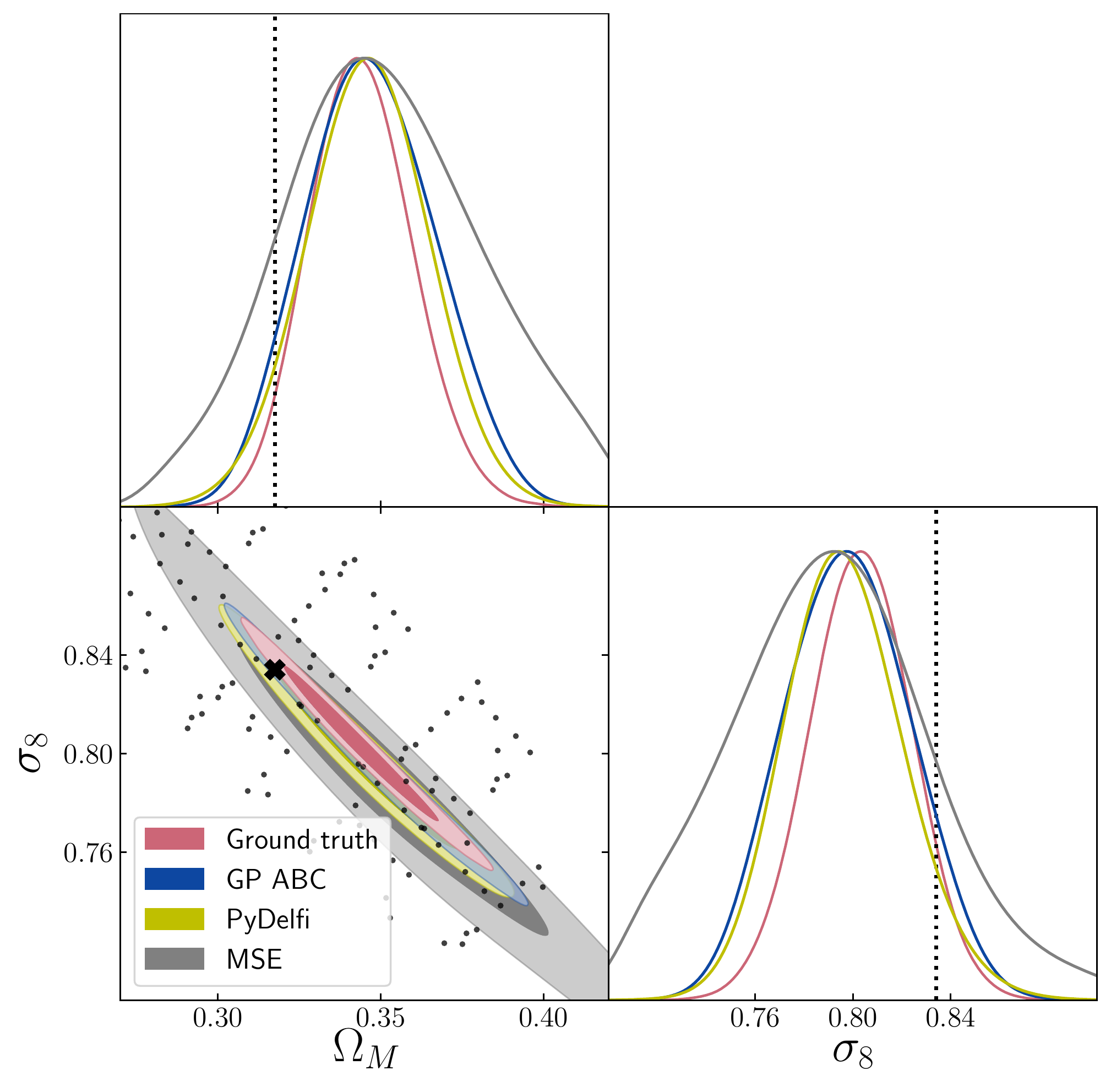}
    \caption{The posterior distributions of the \textbf{2D} model. The fiducial parameter values are indicated with a black cross in the two-dimensional marginal distribution and with vertical dotted lines in the one-dimensional marginal distributions. The black dots correspond to the parameters $\theta_\alpha$ used for the GP ABC, \texttt{PyDelfi} and MSE approach. \label{fig:cons}}
\end{figure}

Using the trained network we perform a GP ABC. We evaluated the ABC posterior estimates as described in section \ref{sec:inference} using a sigmoid kernel
\begin{equation*}
K_h(x) = \frac{2}{\pi h}\frac{1}{\exp\left(\frac{x}{h}\right) + \exp\left(-\frac{x}{h}\right)},
\end{equation*}
with scale parameter $h = 0.35$ for the \textbf{2D} model and $h = 0.45$ for the \textbf{6D} model. We investigate the impact of the scale parameter on the resulting posterior in appendix \ref{ap:scale_param}. For the \textbf{2D} model we use the first 128 parameters of the generated Sobol sequence and fit a Gaussian process using a standard Mat\`ern 5/2 covariance function with automatic relevance determination (ARD). This was done with the \texttt{gpflow} package. We also used the parameter normalization suggested in \cite{Pelle2020}, rotating and scaling the input parameters $\theta_\alpha$ such that they are uncorrelated with zero mean and unit variance. The posterior distribution obtained by running a MCMC with the fitted GP is shown in Figure~\ref{fig:cons}.

For the \textbf{2D} model we performed an additional GP ABC using the predictions of the model trained with the MSE loss using the same parameters, the only difference being that we used the Mahalanobis distance, using the fiducial covariance matrix, as norm. The results are also shown in Figure~\ref{fig:cons}.

For the \textbf{6D} model, we use a Sobol sequence of $n_\theta = 2048$ parameters. In this model, most of the variance is caused by the observational noise. This can be used to effectively increase the sample size $n_x = 50$. For each simulation we create 50 version of observational noise and use the $50\times 50$ mock surveys to estimate the ABC posterior. Afterwards, the same Gaussian process regression as for the \textbf{2D} model is performed. We provide a plot of the two-dimensional marginal posterior distributions in appendix \ref{ap:6d_marginal}.

We use exactly the same data to train a neural density estimator. We use the same model as presented in section 6 of \cite{Alsing2019} only adapting the dimensions. The results can be seen in Figure~\ref{fig:cons}. We also trained a neural density estimator of the summary statistics $\hat{s}$ as opposed to the estimates $\hat{\theta}$ for the \textbf{2D} model. However, the network did not converge to an acceptable result, showing that the distribution of the constructed estimators is easier to learn.

We assess the performance of our method in two ways: by comparing the mean of the posterior and by calculating the Jensen–Shannon divergence (JSD) with respect to the MCMC run on the true posterior. A high precision is extremely important for comsological parameter inference. All results are listed in table \ref{tb:res}. Both methods, GP ABC and \texttt{PyDelfi} perform very well on both models. The GP ABC, however, yields mean parameter values that are on average closer to the ground truth, as well as a lower JSD. The difference of the JSD is larger for the \textbf{2D} model because \texttt{PyDelfi} produces a posterior distribution that is longer in the $S_8 = \sigma_8\sqrt{\Omega_M/0.3}$ direction (i.e. thickness of the degenerate contours). 

\section{Conclusion}
\label{sec:conclusion}
In this work, we presented a novel approach to construct parameter estimators from highly informative summaries. The constructed estimators have a quantifiable bias and are suitable for parameter inference using the presented GP ABC approach or the neural density estimators presented in \cite{Alsing2019}. We validated our approach with a realistic cosmological model, where the constructed estimators are almost unbiased over the whole parameter range while having a close to optimal variance. Given a set of observations, the posterior distribution of the underlying model parameters using GP ABC is close to the true posterior and outperforms the neural density estimators for the \textbf{2D} model and performs equally well for the \textbf{6D} model. Finally, an implementation of the described methods in python is provided on \texttt{github}\footnote{\url{https://github.com/jafluri/cosmo_estimators}}.

\bibliographystyle{apalike_short}
\bibliography{library}

\clearpage

\appendix

 \section{High Dimensional Summary Statistics}
 \label{ap:high_dim}
 
Optimizing the Cram\'er-Rao bound for summary statistics that have a higher dimensionality as the underlying model parameter $d_s > d_\theta$ is not possible with the technique presented in the main paper because the Jacobian is not square and therefore not invertible. However, one can nevertheless define an information maximizing loss, assuming that the Jacobian has a full rank $d_\theta$, we can multiply the information inequality again with the Jacobian
\begin{equation*}
\mathbf{J}^T\mathrm{Cov}_\theta(\hat{s})\mathbf{J} \geq \mathbf{J}^T\mathbf{J}\Im(\theta)^{-1}\mathbf{J}^T\mathbf{J},
\end{equation*}
where we defined 
\begin{equation*}
\mathbf{J} = \frac{\partial \Psi_\theta(\hat{s})}{\partial \theta}.
\end{equation*}
Taking again the log-determinant function we arrive at the following inequality
\begin{align*}
    -\log\det(\Im(\theta)) \leq \ &\log\det(\mathbf{J}^T\mathrm{Cov}_\theta(\hat{s})\mathbf{J}) \\ &- 2\log\det(\mathbf{J}^T\mathbf{J}).
\end{align*}
The right side of this inequality can be used as information maximizing loss for higher dimensional summary statistics and reduces to equation \eqref{eq:loss_1} for $d_s = d_\theta$.

\section{Second Order Estimator}
\label{ap:second_order}

A second order estimator also includes quadratic terms of the summary statistics
\begin{equation}
\hat{\theta}_i(x) = a_i + b_{ij}\hat{s}^j(x) + c_{ijk}\hat{s}^j(x)\hat{s}^k(x). \label{eq:second_oder}
\end{equation}
We start by taking the expected value of the individual terms  of order expansion
\begin{align*}
     \E_{p(x\vert\theta)}\left[a_i\right] &= a_i \\
     \E_{p(x\vert\theta)}\left[b_{ij}\hat{s}^j(x)\right] &= b_{ij}\Psi_\theta^j(\hat{s}) \\
     \E_{p(x\vert\theta)}\left[c_{ijk}\hat{s}^j(x)\hat{s}^k(x)\right] &= c_{ijk}\Psi^j_\theta(\hat{s})\Psi^k_\theta(\hat{s}) \\ 
     &\hspace*{15pt}+ \mathrm{Tr}_{jl}\left[ c_{ijk}\mathrm{Cov}(\theta)^k{}_l \right],
\end{align*}
where $\mathrm{Tr}_{jl}$ is the summation of all terms with $j = l$. Next, we expand all $\theta$ dependent terms around the fiducial parameter up to second order. We start with the expected value
\begin{align*}
    \Psi_\theta(\hat{s})_i =\ & \Psi_{\theta_\mathrm{fid}}(\hat{s})_i \\
                     &+ \frac{\partial\Psi_{\theta_\mathrm{fid}}(\hat{s})_i}{\partial\theta_j}(\theta - \theta_\mathrm{fid})^j \\
                     &+ \frac{\partial\partial\Psi_{\theta_\mathrm{fid}}(\hat{s})_i}{\partial\theta_j\partial\theta_k}(\theta - \theta_\mathrm{fid})^j(\theta - \theta_\mathrm{fid})^k \\
                     &+ \mathcal{O}(\Delta\theta^3).
\end{align*}
We expand the covariance matrix in the same way
\begin{align*}
    \mathrm{Cov}(\theta)_{ij} =\ & \mathrm{Cov}(\theta_\mathrm{fid})_{ij} \\
                     &+ \frac{\partial\mathrm{Cov}(\theta_\mathrm{fid})_{ij}}{\partial\theta_k}(\theta - \theta_\mathrm{fid})^k \\
                     &+ \frac{\partial\partial\mathrm{Cov}(\theta_\mathrm{fid})_{ij}}{\partial\theta_k\partial\theta_l}(\theta - \theta_\mathrm{fid})^k(\theta - \theta_\mathrm{fid})^l \\
                     &+ \mathcal{O}(\Delta\theta^3).
\end{align*}
Plugging these expressions into the expected value of equation \eqref{eq:second_oder} one obtains an expansion of the desired estimator in terms of the statistical moments  of the summary statistics and their derivatives. Assuming that $d_s = d_\theta$ one can obtain $d_\theta^3 + d_\theta^2 + d_\theta$ linear equations, defining $a_i$, $b_{ij}$ and $c_{ijk}$, by demanding that 
\begin{equation*}
    \E_{p(x\vert\theta)}\left[\hat{\theta}_i(x)\right] = \theta + \mathcal{O}(\Delta\theta^3).
\end{equation*}
Solving the set of linear equations is straightforward, however, accurately estimating the higher order moments and their derivative requires a large amount of simulations. Also, it should be noted, that one has to generate additional simulations to obtain perturbations of the fiducial parameter such that one can estimate the higher order derivatives. 

\section{Gaussian Random Fields on the Sphere}
\label{ap:GRF}

Any real function on the sphere $f: S^2 \rightarrow \mathbb{R}$ can be represented as a series
\begin{equation*}
    f(\theta, \phi) = \sum_{\ell = 0}^\infty\sum_{m = -\ell}^\ell \tilde{a}_{\ell m}\tilde{Y}_{\ell m}(\theta, \phi),
\end{equation*}
where the $\tilde{a}_{\ell m} \in \mathbb{R}$ are the real spherical harmonic coefficients and the $\tilde{Y}_{\ell m}(\theta, \phi)$ are the real spherical harmonics which can be expressed in terms of the complex spherical harmonics
\begin{equation*}
    \tilde{Y}_{\ell m}(\theta, \phi) = \left\{ \begin{array}{ll}
        \sqrt{2}\mathcal{R}(Y_{\ell m}(\theta, \phi)) & m > 0 \\
        Y_{\ell 0} & m = 0 \\
        \sqrt{2}\mathcal{I}(Y_{\ell \vert m\vert}(\theta, \phi)) & m < 0 \\
    \end{array} \right.
\end{equation*}
A homogeneous and isotropic Gaussian random field (GRF) on the sphere can be defined as a function where the spherical harmonic coefficients are uncorrelated Gaussian random variables with zero mean
\begin{equation*}
    \tilde{a}_{\ell m} \sim \mathcal{N}(0, \mathcal{C}_\ell),
\end{equation*}
where the variance is given by the (angular) power spectrum $\mathcal{C}_\ell$. The GRF is therefore fully defined by its power spectrum and given a realization of a GRF we can construct an unbiased estimate of the underlying power spectrum via
\begin{equation}
    \hat{\mathcal{C}}_\ell = \frac{1}{2\ell + 1}\sum_{m = -\ell}^\ell \tilde{a}_{\ell m}^2. \label{eq:pseudo_cl}
\end{equation}
It is also useful to define the cross-spectra between two GRF $T_1$ and $T_2$ that might stem from different power spectra, which is usually done in terms if the complex spherical harmonic coefficients 
\begin{equation*}
    \mathcal{C}_{\ell}^{12} = E\left[\mathcal{R}(a_{\ell m}^1)\mathcal{R}(a_{\ell m}^2) + \mathcal{I}(a_{\ell m}^1)\mathcal{I}(a_{\ell m}^2)\right].
\end{equation*}
The cross spectrum can be estimated in the same way as the normal power spectrum and it easy to see that if $\mathcal{C}_\ell^1 = \mathcal{C}_\ell^2$, then, the cross spectrum reduces to the normal power spectrum $\mathcal{C}_\ell^{12} = \mathcal{C}_\ell^1 = \mathcal{C}_\ell^2$.

\begin{figure}[t!]
\includegraphics[width=1.0\linewidth]{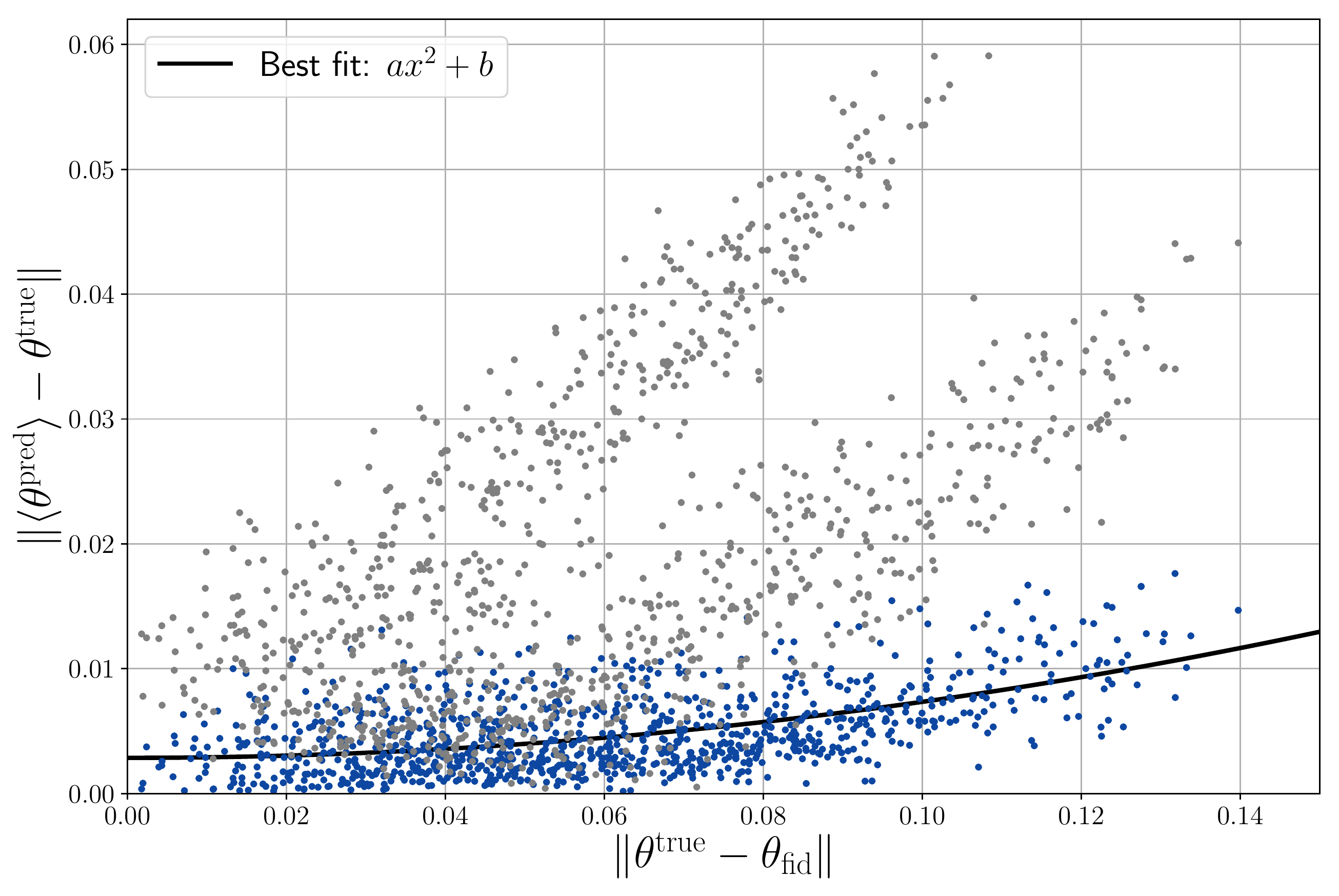}
\caption{Bias of the mean predictions of Figure~\ref{fig:true_vs_pred} plotted against the distance of the underlying parameter to the fiducial parameter. \label{fig:bias}}
\end{figure}

\subsection{Survey Geometry}
\label{ap:suvey_geo}

The $\mathcal{C}_\ell$ estimation described by equation \eqref{eq:pseudo_cl} is only unbiased if the spherical harmonic coefficients were calculated from a full sky map. Realistic surveys, however, do not observe the entire night sky and often have complicated survey masks (see Figure~\ref{fig:zoom}). Such a survey mask leads to a bias in the estimation of the power spectrum called mode-coupling. It is possible to remove this bias with the Wigner matrix formalism. It turns out that the expected value of the estimated masked sky spectrum $\hat{\mathcal{C}}_{\ell,\mathrm{masked}}$ is connected to the underlying power spectrum via a linear transformation
\begin{equation}
    E\left[\hat{\mathcal{C}}_{\ell,\mathrm{masked}}\right]\label{eq:survey_geo} = \sum_{\ell'}W_{\ell\ell'}\mathcal{C}_{\ell'},
\end{equation}
where $W_{\ell\ell'}$ is the Wigner matrix. The Wigner matrix is completely defined by the survey mask and its calculation is described in appendix D of \cite{Sgier2020}.  For continuous GRF it is sufficient to have a non-vanishing area to guarantee that the Wigner matrix is invertible. This holds true most of the time even for pixelized GRF. However, due to numerical stability it is common to apply the Wigner matrix to the underlying power spectrum (theoretical prediction) as oposed to reconstruct an unbiased estimator from the masked power spectrum (observation).

\begin{figure}[t]
    \centering
    \includegraphics[width=0.5\textwidth]{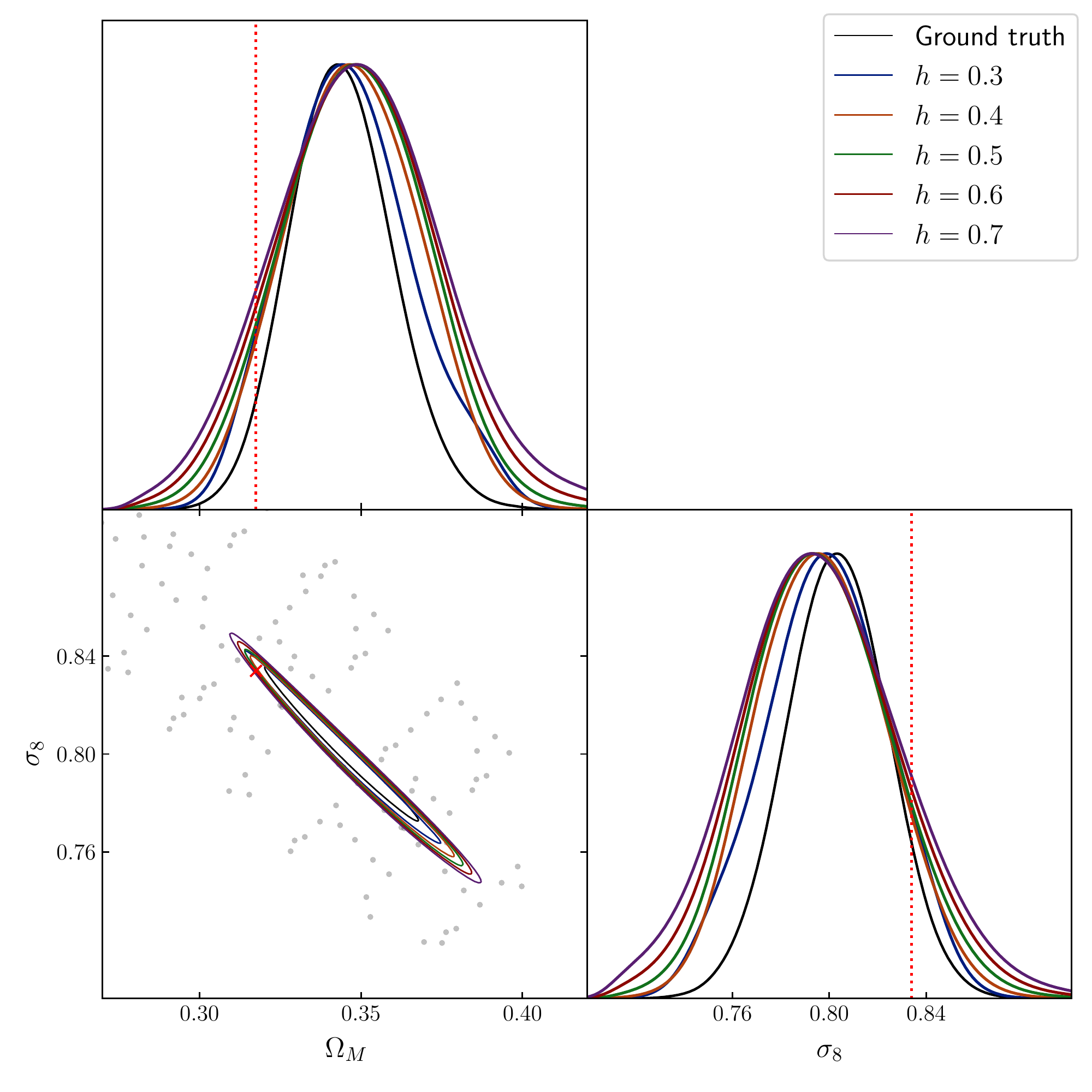}
    \caption{The 68\% confidence contours of the \textbf{2D} model for different scale parameter $h$ compared to the ground truth. \label{fig:scale_dep}}
\end{figure}

\section{Bias of the first Order Estimator}
\label{ap:biases}

The bias of the mean predictions presented in Figure~\ref{fig:true_vs_pred} is shown in Figure~\ref{fig:bias}. The bias is approximately quadratic in the distance to the fiducial parameter, as expected from a first order estimator. The scatter is caused by the finite sample number used to calculate the means and the first order estimator. The shape of the biases of the model trained with the MSE loss is caused by the prior on $S_8$ and are generally larger than those from the first order estimator constructed with the \textbf{2D} model.

\section{Impact of the Scale Parameter}
\label{ap:scale_param}

The estimation of the ABC likelihood requires the setting of a scale parameter $h$ that is used inside the kernel function (e.g. see equation \eqref{eq:ABC_est}). This parameter essentially impacts the ABC posterior as a smoothing factor. A parameter that is chosen too large will lead to a posterior distribution that will be too broad, i.e. smoothed out like a kernel density estimation with inappropriately large bandwidth. If the scale is chosen to small, the posterior estimates will be very noisy because the kernel will map the distance of most simulations to the observation to zero or almost zero. Nevertheless, the posterior distribution of the Gaussian process should not be too sensitve with respect to the scale parameter. We show the posterior of the \textbf{2D} model for differently chosen scale parameters in Figure~\ref{fig:scale_dep}. As expected, the contours become broader as the scale parameter increases. All of the generated posterior distributions are consistent with the ground truth. 

\section{Marginal Distributions of the \textbf{6D} Model}
\label{ap:6d_marginal}

\begin{figure*}
	\centering
	\includegraphics[width=0.9\linewidth]{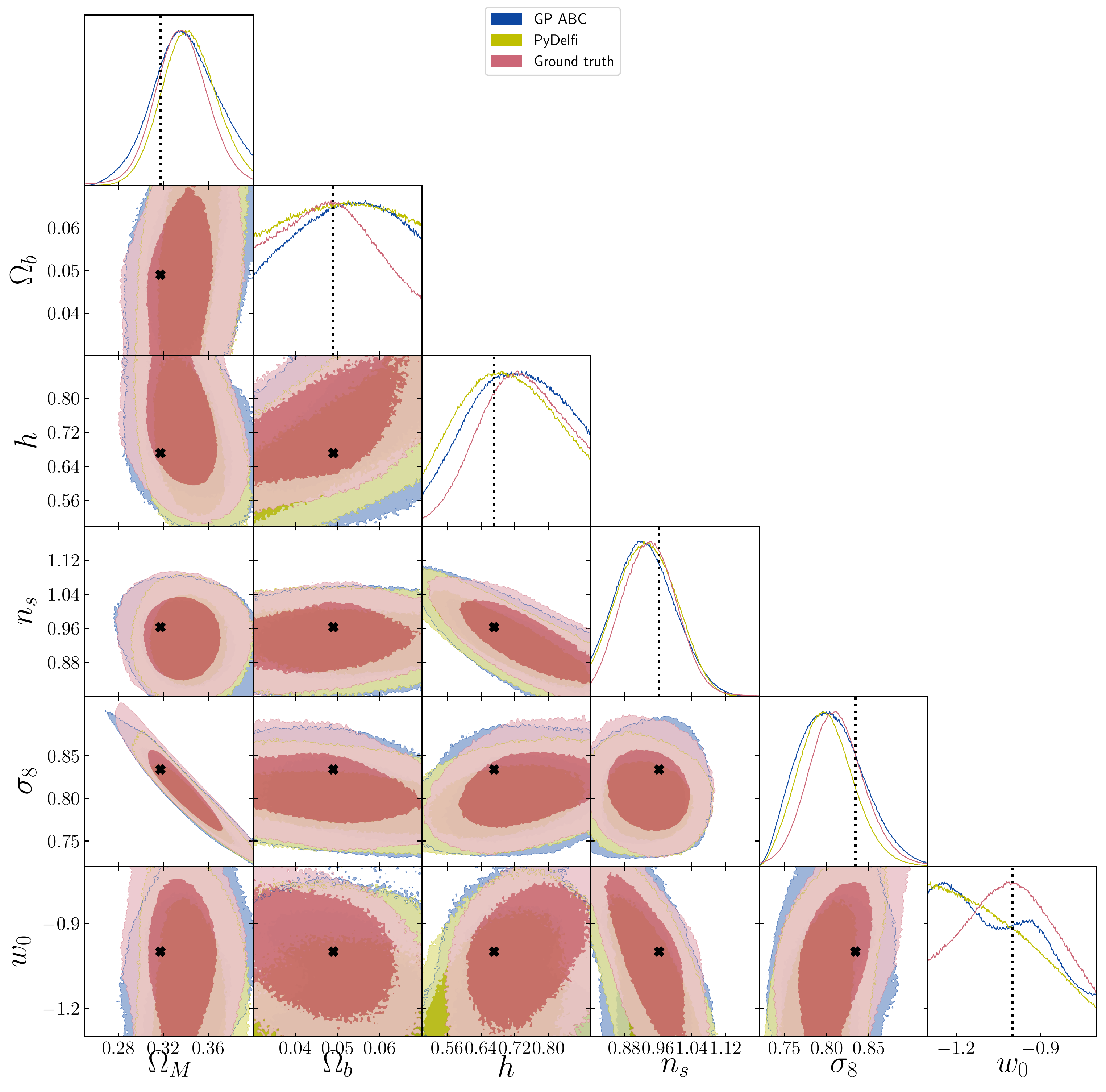}
	\caption{The posterior distributions of our \textbf{6D} model. The fiducial parameter is indicated with a black cross in the two-dimensional marginal distributions and with vertical dotted lines in the one-dimensional marginal distributions. \label{fig:cons_6d}}
\end{figure*}

We show a plot of the marginal distribution for our \textbf{6D} model obtained with GP ABC and neural density estimators in Figure~\ref{fig:cons_6d}. The marignal distribution from both methods are in excellent agreement with the true posterior. 

\end{document}